# Young's double-slit, invisible objects and the role of noise in an optical epsilon-near-zero experiment

*Daniel Ploss[1,‡], Arian Kriesch[1,‡], Christoph Etrich[2], Nader Engheta[3], and Ulf Peschel[2,*]*

[1]Institute of Optics, Information and Photonics and Erlangen Graduate School in Advanced Optical Technologies (SAOT), Friedrich-Alexander University Erlangen-Nürnberg (FAU), Haberstr. 9a, Erlangen 91058, Germany

[2]Institute of Condensed Matter Theory and Optics, Friedrich-Schiller-University Jena, Max-Wien-Platz 1, Jena 07743, Germany

[3]Department of Electrical and Systems Engineering, University of Pennsylvania, 200 South 33rd Street, Philadelphia, Pennsylvania 19104-6314, USA

ABSTRACT

Epsilon-near-zero (ENZ) media disclose the peculiarities of electrodynamics in the limit of infinite wavelength but non-zero frequency for experiments and applications. Theory suggests that wave interaction with obstacles and disturbances dramatically changes in this domain. To investigate the optics of those effects we fabricated a nanostructured 2D optical ENZ multilayer waveguide that is probed with wavelength-tuned laser light via a nanoscale wave launch configuration. In this experimental framework we directly optically measure wave propagation and diffraction in a realistic system with the level and scale of imperfection that is typical in nanooptics. As we scan the wavelength from 1.0 $\mu$m to 1.7 $\mu$m, we approach the ENZ regime, and observe the interference pattern of a micro-scale Young's double slit to steeply diverge. By evaluating multiple diffraction orders we experimentally determine the effective refractive index $n_{eff}$ and its zero-crossing as an intrinsic measured reference, which is in agreement with theoretical predictions. We further verify that the double-slit and specifically placed scattering objects become gradually invisible when approaching the ENZ regime. We also observe that light-matter-interaction intensifies towards ENZ and quantify how speckle noise, caused by tiny random imperfections, increasingly dominates the optical response and blue-shifts the cut-off frequency.

MAIN TEXT

Light is an electromagnetic wave, characterized by its frequency and wavelength, where the former mostly determines its interaction with matter and the latter defines its diffraction. Usually both quantities are closely linked to each other by the refractive index of the surrounding material. Only since the introduction of metamaterials this strict relation has become widely tailorable. The extreme condition that we investigate in this letter recently attracted particular interest, namely the relative permittivity ε of the surrounding medium approaching zero (epsilon-near-zero, ENZ), which causes the wavelength to diverge for a given finite frequency[1,2].

The fact that in those epsilon-near-zero (ENZ) media geometric dimensions of any object placed in the beam path become tiny compared to the local wavelength leads to a variety of interesting effects such as supercoupling through small channels[3], enhancement of nonlinear effects[4] and of elevated photon density states for embedded emitters[5]. As light-matter interaction is highly affected in the ENZ regime the field evolution is considerably modified.

Here, we conduct several experiments on wave propagation and diffraction in a two-dimensionally extended ENZ environment that we excite and probe optically. Thus we test and measure quantitatively the specific ENZ effects caused by modified wave interaction with single and multiple obstacles, and double slits, but we also find some unexpected features[6], particularly noise levels that rise as we scan the frequency towards ENZ.

We transfer one of the most celebrated investigations on wave propagation - Young's double-slit experiment - into an ENZ environment. Thomas Young showed in 1807 that light waves after simultaneously passing two adjacent slits form a wavelength-dependent interference pattern, thus demonstrating the wave nature of light[7]. Young's double-slit experiment was later used to prove the wave character of matter, as for fundamental particles e.g., electrons[8] and neutrons[9], and quasi particles like plasmons[10,11].

We observe that Young's interference pattern steeply diverges (see Figs.1a and 1c) when scanning the frequency towards the ENZ regime. We determine based on this measurement the frequency-dependent permittivity $\varepsilon(\omega)$, which serves as a direct reference that is robust against systematic and fabrication errors for other diffraction experiments on the same sample.

We further measure in this ENZ environment how diffraction caused by artificial obstacles that are specifically placed in the beam path successively vanishes (Figs. 1b and 1d) when approaching the ENZ regime, proving theoretical claims[2].

Artificial ENZ materials for optical frequencies can be produced in various ways: 1. With transparent conductive oxides (TCOs)[12,13] by tuning their charge-carrier density, which was recently applied for an electro-optical modulator[14–16] and which in the ENZ domain give even rise to enhanced nonlinearity[17]. 2. As hybrid plasmonic waveguides[18] that close to cut-off may act as photonic wires for optical "metatronic" nanocircuitry[19,20]. 3. As metamaterials, pioneered by microwave experiments[2], realized in the optical domain with stack[21] geometries, which are usually lossy, with embedded metal cubes[22] or pillar[23] geometries, or as dielectric/semiconductor heterostructuctures[24], all of which can only be realized as a thin layer and are hence not useful as an extended ENZ medium for conducting an optical diffraction experiment inside the medium. Our study is inspired by a metal-clad planar rectangular dielectric geometry that was recently introduced in form of a few 100 nm small entirely closed resonator whose photonic modes close to cut-off were externally excited by an electron beam[25].

We demonstrate an optically excited two-dimensionally extended ENZ waveguide where light propagation is confined to an insulator sandwiched between two metal layers (see Figs. 1e and 1f). In these 2D waveguides we implement Young's double slit experiment, place specific obstacles and combinations side by side and probe and characterize both types of systems in the same, reproducible ENZ regime.

Metal-insulator-metal (MIM) waveguides are produced by sequentially sputtering layers of silver, silica and silver again onto a glass substrate thus forming sandwiched dielectric waveguides. We fabricate two different waveguide thicknesses of 440 nm and 535 nm resulting in an analytically predicted cut-off of the lowest order TE-mode (E-field polarized parallel to the metal, mainly in y-direction, Figs. 2a,b) at $\lambda_1^{ana} = 1.40\ \mu$m and $\lambda_2^{ana} = 1.67\ \mu$m, respectively. Experiments are performed in a spectral range that includes this cut-off. As the plasmonic lowest order TM waveguide mode (E-field predominantly polarized normal to the metal in x-direction, Figs. 2a,b) shows no cut-off, a well-defined and polarization-specific excitation is realized by a stripe waveguide being connected to the 2D ENZ waveguide on one side (i.e., the "input" side) and cut open under a 54° angle on the other side (i.e., the "output" side) for external excitation and detection. A linearly polarized Gaussian beam is focused onto this input facet from above (see Figs. 1e and 1f). The light is transferred to the ENZ waveguide while maintaining a well-defined polarization state. After passing through the ENZ waveguide the field at the output is monitored by imaging a 50 $\mu$m slice of light irradiated by an oblique y-directed cut through the whole MIM waveguide acting as a resonant out-coupling "antenna" (Fig. 2a) (for more details on the set-up[26,27] and fabrication see supporting information).

To implement Young's experiment in such ENZ environment we integrate two slits of varying separations, i.e., pitch $p$, into the MIM waveguides (Figs. 2a,b), and observe the interference patterns at the output. Well-defined patterns up to the ± 6$^{th}$ interference order (for $p = 8$ $\mu$m) are detected at the out-coupling facet. As the operating wavelength increases, the stripes in the patterns expand and a decreasing number of interference orders stays in our field of view. Close to the ENZ condition, i.e., near the cut-off of the TE mode of the waveguide, only the zeroth interference order remains and finally the detected field becomes almost homogeneous, thus rendering the original structure undetectable, i.e., invisible.

To gain further confirmation, our set-up is also utilized to perform sensitive measurements of the effective refractive index of the waveguide mode by evaluating the wavelength-dependent positions of the interference orders. Using the well-known double-slit diffraction formulation (including a non-paraxial correction, see supporting information) we fit the positions of the experimentally determined interference maxima (see the white lines in Figs. 2c-f) and thus determine the effective refractive index $n_{eff}$ of the TE mode as a function of the free-space wavelength. The spectral range of our index measurement is only limited by the required presence of at least the first interference order. For the two waveguides of different thickness $d_1 = 440$ nm and $d_2 = 535$ nm, the interference pattern disappears from the field of view around a wavelength of $1.25\,\mu$m and $1.49\,\mu$m (see Figs. 2 g,h), respectively, resulting in the experimentally determined minimum effective index of $n_{min\_1} = 0.28 \pm 0.07$ and $n_{min\_2} = 0.22 \pm 0.07$. By extrapolating the measured effective index values we experimentally determine the position of the cut-off as $\lambda_1^{exp} = 1.29\,\mu\text{m} \pm 0.04\,\mu\text{m}$ and $\lambda_2^{exp} = 1.51\,\mu\text{m} \pm 0.08\,\mu\text{m}$, respectively (see Figs. 2g,h). These values coincide well with a breakdown of TE transmission for the whole ENZ waveguide (see Figs. 4b,c and supporting information).

Following our findings on the ENZ version of the Young double-slit problem, we now turn our attention to another diffraction scenario, namely an obstacle placed in the beam path, in order to examine a peculiarity of scattering in the ENZ environment. Indeed we observe analogous drastic modification of the diffraction patterns of the object, as the operating wavelength approaches the cut-off frequency of the TE mode. Vertically focused ion beam (FIB) drilled holes of $0.5\,\mu$m diameter (see Figs. 3a-d), which we added to the MIM waveguides $21\,\mu$m behind the input antenna are found to cast a dark shadow surrounded by a weak diffraction pattern both in our simulations and experiments (see Figs. 3f,i and Figs. 3g,j, for the corresponding transmission plots see supporting information). As the operating wavelength is increased the diffraction around these obstacles becomes prominent and a bright spot appears in the center of the shadow, demonstrating the first observation of the Poisson spot[28] in an ENZ structure (see Fig. 3f). As numerical simulations confirm this Poisson spot diverges in the ENZ regime, rendering all traces of the obstacle in the beam path experimentally undetectable (see Figs. 3i,j).

It is worth mentioning that owing to structural imperfections of our samples, experimentally determined field distributions show noticeable deviations from the ideal behavior. A thorough investigation of FIB cuts reveals that although the waveguide layers have the accurate thicknesses their interfaces possess a residual corrugation caused by the material with a vertical depth of about 60 nm and a subwavelength transverse correlation length of about 250 nm (see Figs. 2b,i and Figs. 3b,d). While being negligible in the short wavelength range, scattering from these defects and resulting speckles grow significantly around the experimentally determined cut-off wavelengths leading to an accumulation of noise inside the waveguide, both in experiments and simulations (see Figs. 4a,d and for more information regarding investigations on speckle noise see supporting information). Moreover, as imperfections break the translational symmetry of the film waveguide scattering causes a conversion of TE polarized light into the orthogonal polarization (TM) state - a process, which is again significant close to the TE-cut-off only (see Figs. 4b,c and supporting information Fig. S13).

Arguably the most unexpected effect of the structural imperfections is a significant shift of the cut-off wavelengths. An eigenmode analysis predicts analytical cut-off wavelengths of $\lambda_1^{ana} = 1.40\,\mu$m for the thin, respectively $\lambda_2^{ana} = 1.67\,\mu$m for the thicker waveguides. But, all experimental results including refractive effective index measurements (Figs. 2g,h), the transmission curves (Fig. 4b) and a comparison between diffraction measurements and simulations (Figs. 3f,g,i,j) independently point to a cut-off wavelength of the TE mode of $\lambda_1^{exp} \approx 1.29\,\mu$m and $\lambda_2^{exp} \approx 1.55\,\mu$m, respectively. This noticeable blue shift is larger than our measurement error (Figs. 2g,h). TE-polarized light in the ENZ domain seems to be influenced by the corrugated structure of the interfaces thus experiencing an effectively reduced waveguide thickness and therefore a blue-shifted cut-off wavelength. In fact, the measured TE cut-off wavelengths correspond to those of unperturbed waveguides with a considerably reduced thickness of 400 nm and 480 nm instead of the original values of 440 nm and 535 nm, respectively. Numerical simulations taking into account random imperfections (Figs. 4d,e,f) demonstrate that those are responsible for a significant reduction of transmission and for the observed cut-off shift.

In conclusion, Young's double-slit experiment was conducted in an ENZ environment. The interference patterns steeply, but predictably diverge as the ENZ regime is approached. From the interference patterns we extract the wavelength-dependent effective refractive index $n_{eff}$ and derive the spectral positions of the vanishing effective permittivity ε. This provides an intrinsic, precise reference for further quantitative diffraction experiments. Close to this frequency clear patterns from light interaction with single and double assemblies of obstacles placed in the light path successively wash out until they disappear. Hence, the double slit as well as any obstacle placed in the beam path become virtually invisible in the ENZ regime. As the effective refractive index tends to zero, light-matter interaction is intensified and backscattering from tiny random imperfections induces a significant blue shift of the cut-off frequency. In addition the enhanced coupling between light and structural imperfections also results in growing levels of speckle noise close to cut-off, which finally dominate all features.

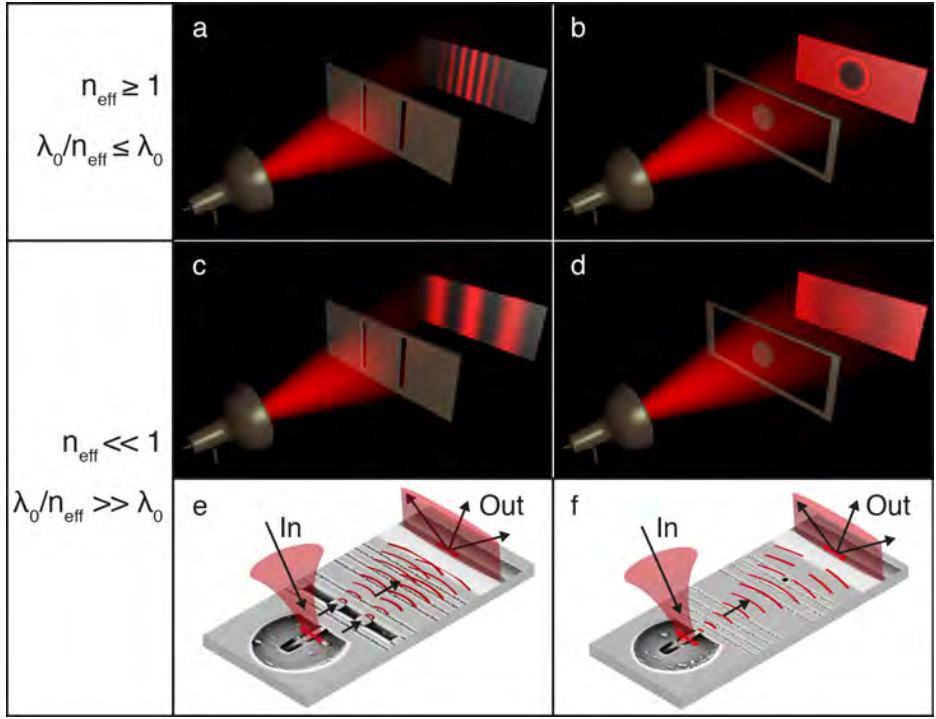

**Figure 1.** Young's double-slit experiment (a and c) and wave diffraction from an obstacle (b and d) are heavily influenced by the effective index $n_{eff}$ and the effective wavelength in the structure. In our experiments light propagates in 2D metal-insulator-metal waveguides where the effective index of the lowest order TE mode (electric field parallel to the metal, see red arrows in (e,f)) vanishes at cut-off frequency. We demonstrate that (a,c) the double slit diffraction pattern expands and that (b,d) the optical impact of obstacles diminishes as we approach the ENZ regime.

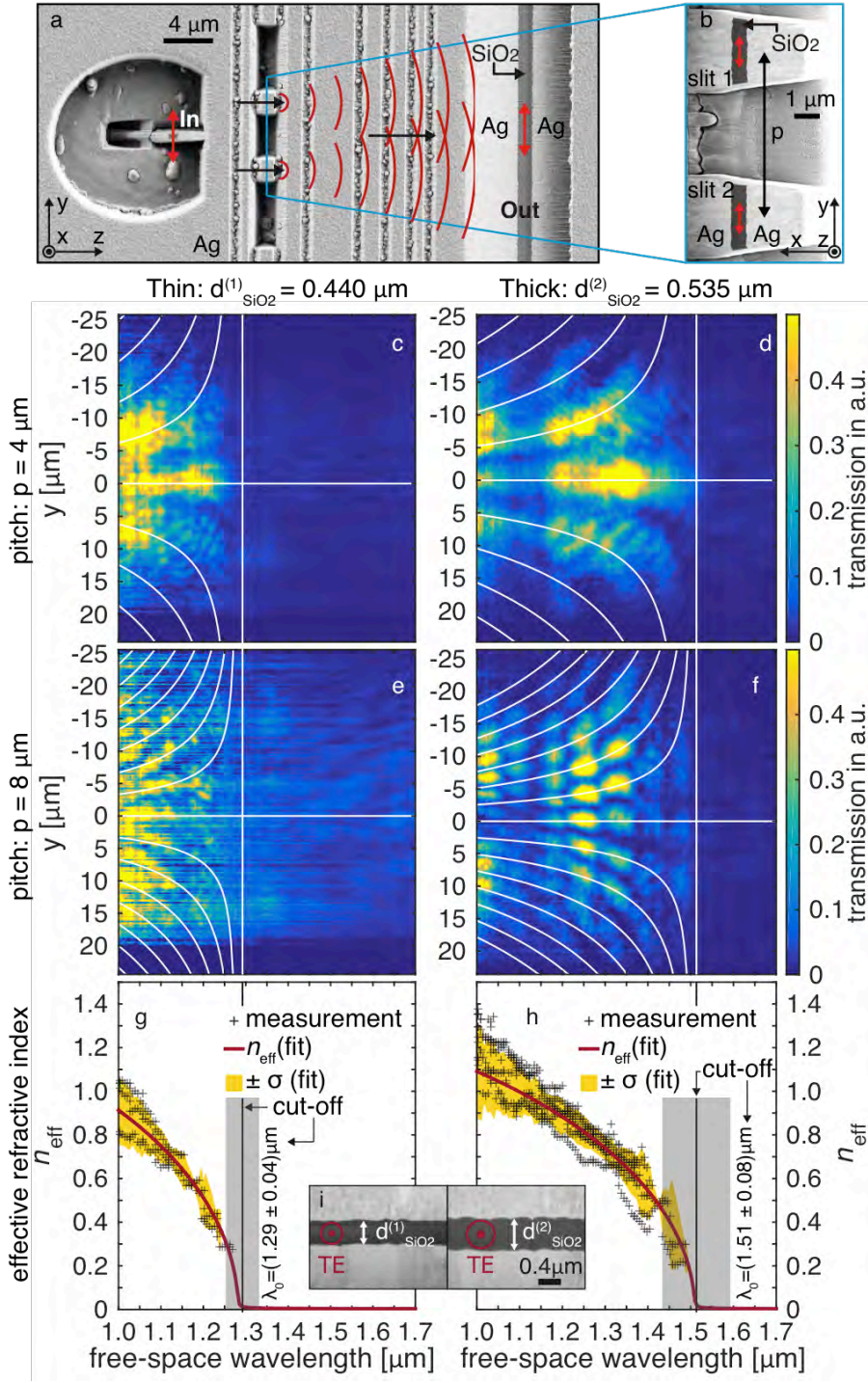

**Figure 2.** Young's double-slit experiment conducted in a waveguide-based ENZ environment. (a) SEM image of the sample and (b) the enlarged view of the double-slit cross-section. On the top metal surface (a) an aperiodic grating scatters out small residual surface plasmon polaritons that would reduce the SNR when reaching the out-coupling slit. (c-f) Experimentally recorded

TE-polarized diffraction patterns for different waveguide thicknesses and pitches compared with a best fit (white lines) based on an estimated $n_{\text{eff}}(\lambda_0)$. The excitation is always in TE-polarization (y-direction). (g,h) Effective refractive indices of the waveguide determined by fitting double slit diffraction patterns (i) SEM cross-section image for the two different waveguide thicknesses (equally scaled).

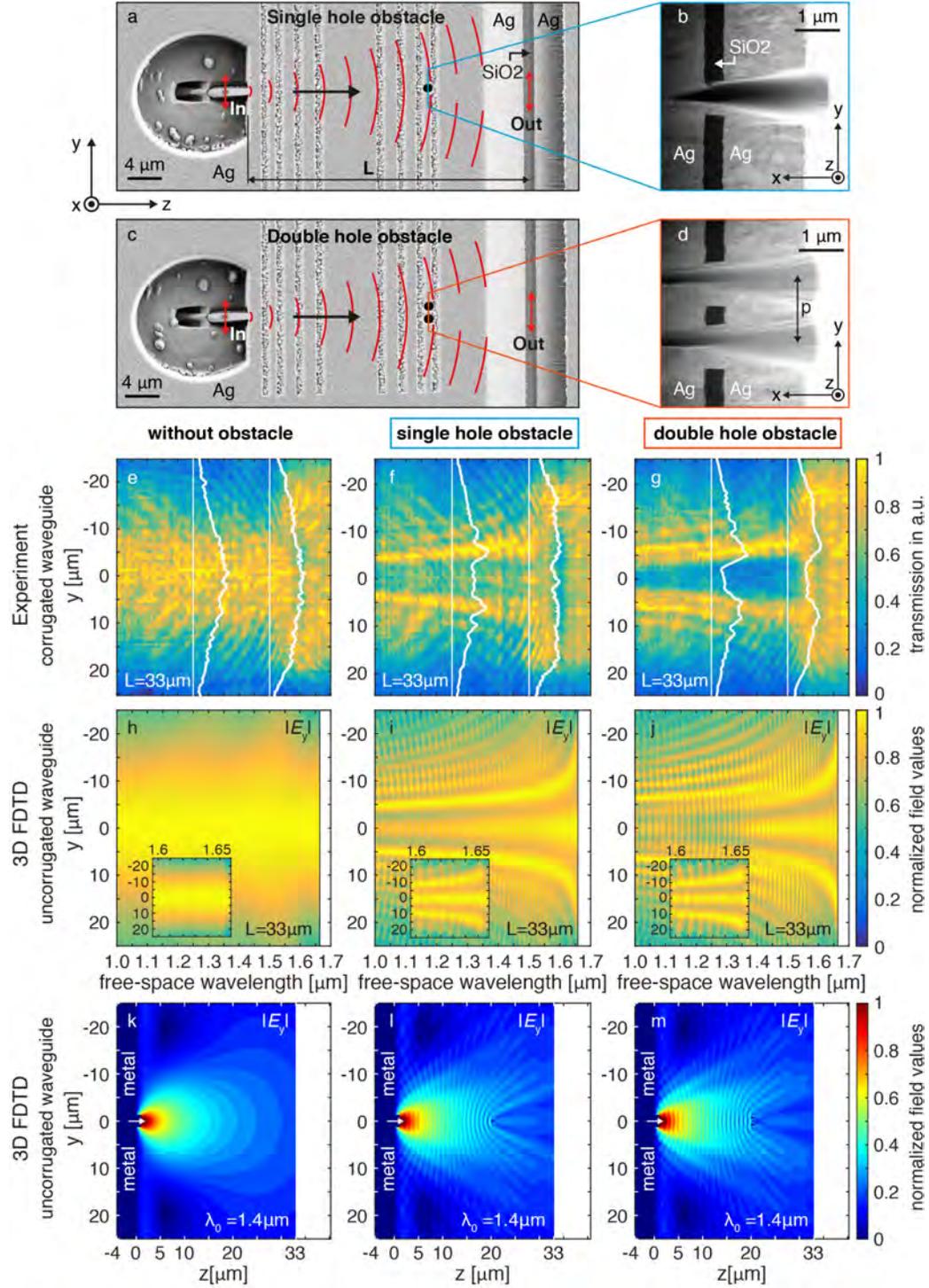

**Figure 3.** Fields in the ENZ waveguides with $d_2 = 0.535$ $\mu$m. (a,c) SEM images of the structures with opaque obstacles in form of a metallized single hole and double holes drilled into the guiding layer and (b,d) an enlarged view of the cross-section of these single/double holes. (e-g)

Measured (with binned cross-sections from $\lambda_0 = 1.25\,\mu$m - $\lambda_0 = 1.32\,\mu$m and from $\lambda_0 = 1.50\,\mu$m - $\lambda_0 = 1.57\,\mu$m) and (e,h) in the waveguide with no obstacle (f,i) behind the single hole and (g,j) behind the double hole obstacle (p = 1.5 $\mu$m). (k-m) Simulated field profiles show a dipole like emission from the connected stripe waveguide (k) inside the ENZ waveguide, (l) in the presence of the single hole and (m) the double hole.

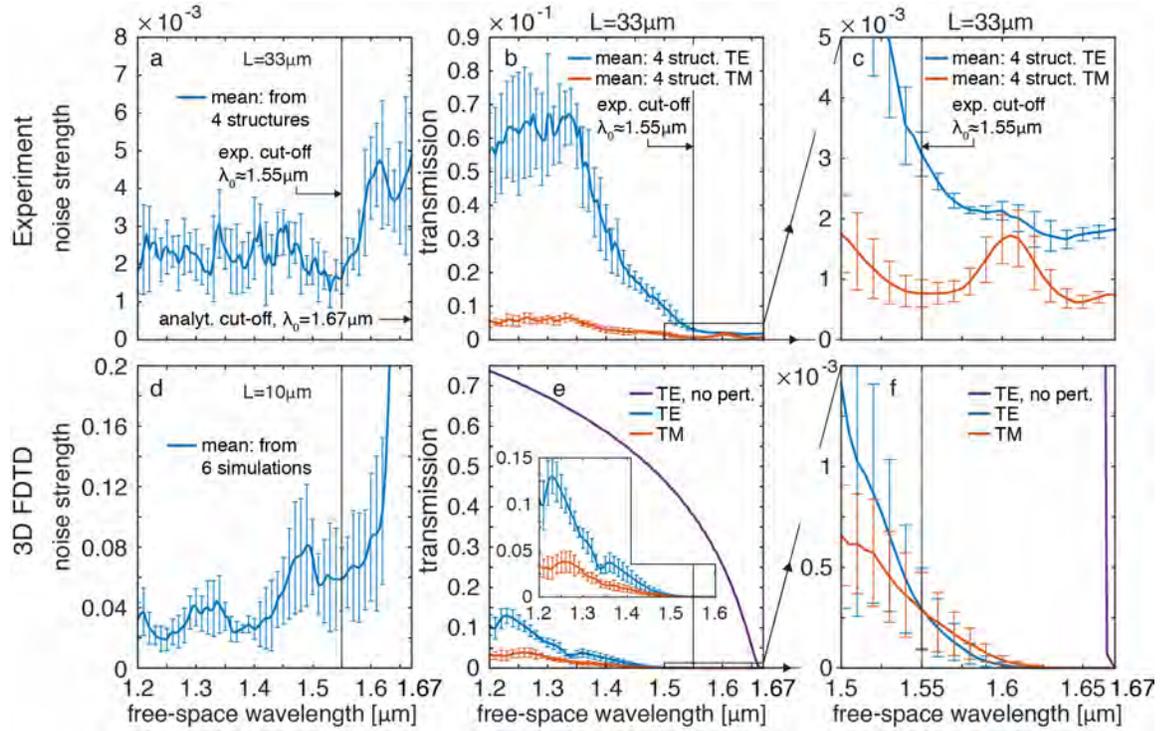

**Figure 4.** The effect of imperfections in the ENZ waveguides ($d_2 = 535$ nm). (a,d) A considerable growth of noise is observed in the ENZ regime, both in (a) experiment and (d) simulation (See supporting information on calculating the noise strength in (a,d)). (b,e) Transmission (b) measurements and (e) simulations (no perturbation in the waveguide: purple curve) display a breakdown of TE transmission at the experimentally determined cut-off ($\lambda_2 \approx 1.55\,\mu$m). Between the experimental $\lambda_2^{exp} \approx 1.55\,\mu$m and the analytical $\lambda_2^{ana} \approx 1.67\,\mu$m cut-off still a weak, but noticeable transmission with strong TM content is detected (c)

experimentally and less intense (f) numerically. The results were averaged over 4 experimental structures and 6 numerical realizations. The excitation is always TE-polarized.

METHODS

**Sample fabrication and structuring**

MIM waveguide: A silica quartz glass with a thickness of 170 $\mu$m was used as substrate. The MIM waveguide layers were fabricated with a first PVD deposition (magnetron sputtering with an AJA ATC-Orion-Series machine) of four layers in total (50 nm Cr, 1.5 $\mu$m Ag, 440 nm and 535 nm pure $SiO_2$, 1.5 $\mu$m Ag, see Fig. S1).

In-coupling antenna for excitation: The in-coupling-antenna configuration, which was optimized with FDTD simulations, was created with focused ion beam (FIB) milling (Zeiss-NVision 40 crossbeam) to achieve an almost rectangular cross section of the dielectric waveguide channel (in-coupling antenna) of 420 nm or 460 nm in width and 440 nm or 535 nm in height. A second PVD deposition of two layers (800 nm Ag, 50 nm Cr) creates the metal sidewall on both sides of the in-coupling antenna (see Fig. S2 in the supporting information), followed by a necessary sidewall FIB milling in order to get straight walls with a metal thickness of 200 nm on each side, which results in a one dimensional rectangular closed MIM waveguide (in-coupling antenna). Finally the in-coupling antenna was structured again with FIB in order to have an oblique excitation antenna configuration with an angle of $\phi = 54°$, with respect to the surface normal. The angle was chosen such that a perpendicular incident focused laser beam excites the in-coupling antenna from the far field. The light enters the dielectric channel. The in-coupling antenna emits a clean mode into the two dimensional MIM waveguide, which has effectively infinite size in y-direction.

Out-coupling antenna for probing: At the end of the MIM waveguide the out-coupling antenna was structured using FIB with the same oblique cutting angle as used for the in-coupling antenna. The resulting long oblique slit in y-direction through all layers of the waveguide was designed (optimized with FDTD simulations) to scatter out the light along its extension, such that it can be clearly imaged in the far field with the experimental setup.

Young's double slit, single and double hole: Young's double slit was realized with three different pitches ($p = 4, 6, 8\,\mu$m) by means of FIB cutting into the two dimensional waveguide (see Figs. S3, S4 in the supporting information). A single hole (top ø = 1 $\mu$m, layer ø ≈ 0.4 $\mu$m) and a double hole (pitch $p = 1.5\,\mu$m, open layer slit between holes $s \approx 0.9\,\mu$m) as scattering objects were structured with the same procedure.

**Optical measurement of the fabricated samples**

Samples were placed on a 3D-piezo stage (Physik Instrumente (PI) GmbH & Co. KG) in a custom made optical scanning far-field setup (see details in the supporting information, see Fig. S7). The collimated beam from a supercontinuum light source (SuperK Extreme EXR-15, NKT Photonics, $\lambda_0 = 0.48\,\mu$m – 2.4 $\mu$m) is spectrally filtered by a programmed acousto-optic spectral tunable filter (AOTF, operated at $\lambda_0 = 1.0\,\mu$m – 1.7 $\mu$m, resolution accuracy of $\Delta\lambda_0 = 5$ nm fwhm) and is then coupled into a single mode fiber SMF28. This fiber filters a sufficiently pure fundamental Gaussian mode shape for the most sensitive measurements, depending on absolute power transmission in the range of $\lambda_0 = 1.2\,\mu$m – 1.7 $\mu$m and for less sensitive reference measurements in an extended range.

Light is coupled out of the fiber, collimated and enters the experimental setup through a linear polarization filter (polarizer 1) and a non-polarizing beam splitter (NPBS) that directs 50 % of the power to a reference diode (InGaAs). The main beam (50 % of in-coupling power) enters a high NA microscope objective (Leica, HCX PL Fluotar 100x/0.90) and is focused to a diffraction-limited spot (ø < 2 $\mu$m) that is adjusted with the piezos to cover the in-coupling antenna of the ENZ structure. Using the vertical displacement of the 3D-piezo stage, the focus is adjusted on the in-coupling antenna. Light that is propagated through the waveguide couples out and is then collected with the same objective. A NPBS reflects this imaging beam out of the excitation beam path and towards a second linear polarization filter (polarizer 2). For cross-polarization measurements polarizer 2 is set perpendicular to polarizer 1 with a maximum suppression ratio of 1:10000. After polarizer 2, light passes the tubus lens to form a real image on an InGaAs NIR CCD camera (Xenics XS, 320 x 256 pixels, pixel pitch = 30 $\mu$m, 14 bit). The in-coupling polarization can be changed with polarizer 1 for selectively exciting the photonic TE mode or the plasmonic TM mode inside the MIM waveguide. The rectangular shaped in-coupling antenna supports both modes. The out-coupled light can be filtered with polarizer 2 in order to selectively detect the parallel- or cross-polarized component of the electromagnetic field that is emitted.

**Evaluation of the effective refractive index of the propagating TE mode**

The resulting intensity distribution of Young's double-slit experiment in the transversal *y*-direction detected with the IR CCD camera follows

$$I(y) = \frac{I_0}{2} \cdot \frac{\sin^2\left(\frac{\pi \cdot D}{\lambda_{\text{eff}} \cdot L} \cdot y\right)}{\left(\frac{\pi \cdot D}{\lambda_{\text{eff}} \cdot L} \cdot y\right)^2} \cdot \left(1 + \cos\left(\frac{2\pi \cdot p}{\lambda_{\text{eff}} \cdot L} \cdot y\right)\right),$$

with *D* being the single slit width, *p* the pitch of the double-slit, *L* the distance between the double slit and the out-coupling antenna and $\lambda_{\text{eff}}$ the effective wavelength inside the MIM waveguide. The cos-part represents the interference pattern with respect to *p* and *L*, and the sinc-part describes the final envelope, due to diffraction on each single slit with the slit width *D*. This formula is only valid in the paraxial limit, e.g., for diffraction angles $\phi \leq 22°$. In the experiment the maximum captured orders refer to a diffraction angle above $\phi = 40°$ exceeding the diffraction angle for the paraxial limit by almost the factor of two. Hence, the paraxial error $\Delta_{\text{ParError}}(y) = y - b(y)$, with

$$b(y) = \tan^{-1}\left(\frac{y}{L}\right) \cdot L,$$

was corrected for each measured image before starting with the image evaluation for the effective refractive index of the propagating mode. To start with, a Fourier transform of the measured intensities over the lateral coordinate $y$ was performed in order to translate the transverse distances between the resulting intensity minima and maxima of the interference pattern into frequency space [$2\pi/y$]. This is done for the whole interference result (line wise in $y$ direction) for each measured free space wavelength, i.e., for the total spectral range of $\lambda_0 = 1.0\,\mu\text{m} - 1.7\,\mu\text{m}$ in steps of $\Delta\lambda_0 = 5$ nm for each of the resulting 141 wavelength points (= 700 nm/5 nm + 1 steps). The resulting frequency was inserted into the cos-part of Young's intensity distribution $I(y)$ in order to calculate the effective wavelength $\lambda_{\text{eff}}$ and from that the effective refractive index $n_{\text{eff}}$. Details for the fit of all resulting $n_{\text{eff}}$ in order to get the final effective refractive index are explained in the supporting information.

**3D FDTD Simulations**

The commercial software package Lumerical 3D FDTD Solution® was used in order to simulate the fabricated metal-insulator-metal waveguide, including the "rough", "corrugated" structure in between both sides of metal and glass. Only parts of the fabricated structure with a reduced size compared to experiments could be numerically simulated due to limited computational resources. A more computationally efficient, but less flexible in-house-developed code for 3D FDTD written in Fortran with no incorporated roughness in the waveguide was used with the aim to cover the large-scale wave propagation through the ENZ structure with identical dimensions as in the experiment (see more details in the supporting information).

## ASSOCIATED CONTENT

**Supporting Information**.

Detailed descriptions on the applied fabrication, measurement, evaluation, theory and simulation techniques. The supporting information contains Figures S1 – S27.

## AUTHOR INFORMATION

**Corresponding Author**

\* E-mail: ulf.peschel@uni-jena.de.

**Author Contributions**

D.P., A.K. and U.P. conceived the idea for the experiment, the structure design and evaluated the results. D.P. fabricated the samples and performed the optical measurements. D.P. and C.E. performed the numerical simulations. The project was under the supervision of N.E. and U.P. The manuscript was written with contributions from all authors. All authors have given approval to the final version of the manuscript. ‡D.P. and A.K. contributed equally to this work.

## ACKNOWLEDGMENT

We thank the Max-Planck-Institute for the Science of Light (MPL) in Erlangen, Germany for providing us with the production facilities (PVD, SEM, FIB). D. Ploss and A. Kriesch acknowledge funding from the Erlangen Graduate School in Advanced Optical Technologies (SAOT) in the framework of the German excellence initiative. A. Kriesch acknowledges funding by the Cluster of Excellence Engineering of Advanced Materials (EAM), Erlangen, Germany.

TABLE OF CONTENTS FIGURE

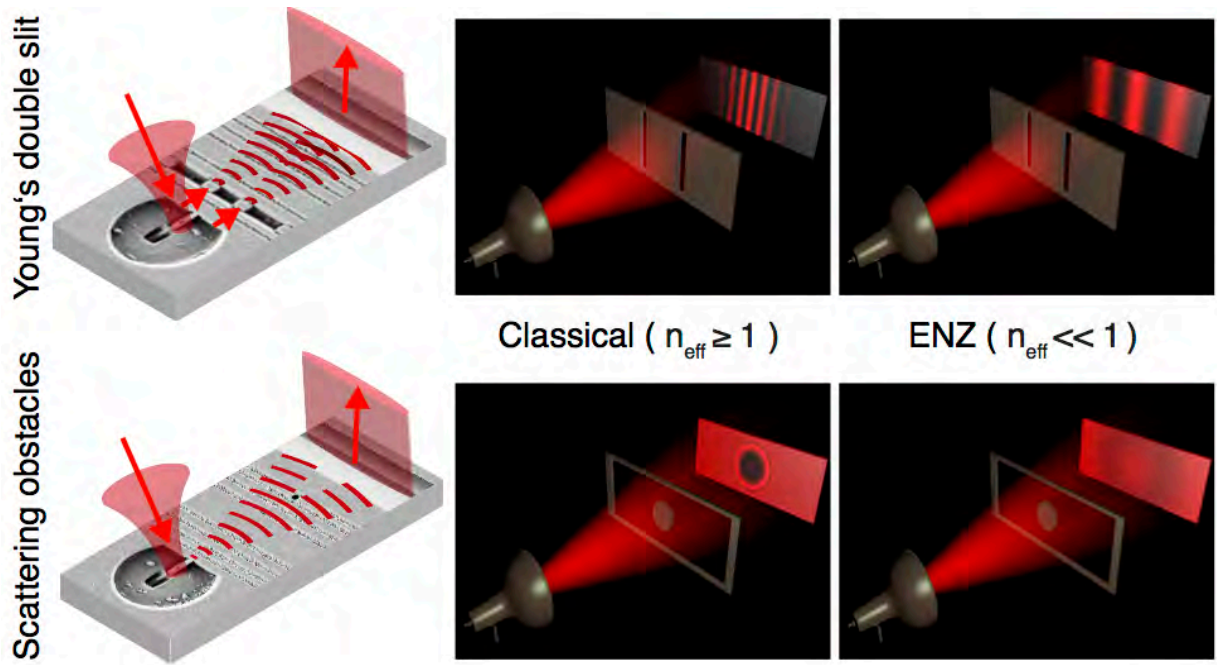

Supporting Information

# Young's double-slit, invisible objects and the role of noise in an optical epsilon-near-zero experiment

**Authors:**

*Daniel Ploss[1,‡], Arian Kriesch[1,‡], Christoph Etrich[2], Nader Engheta[3], and Ulf Peschel[2,†]*

**Affiliations:**

[1]Institute of Optics, Information and Photonics and Erlangen Graduate School in Advanced Optical Technologies (SAOT), Friedrich-Alexander University Erlangen-Nürnberg (FAU), Haberstr. 9a, Erlangen 91058, Germany

[2]Institute of Condensed Matter Theory and Optics, Friedrich-Schiller-University Jena, Max-Wien-Platz 1, Jena 07743, Germany

[3]Department of Electrical and Systems Engineering, University of Pennsylvania, 200 South 33rd Street, Philadelphia, Pennsylvania 19104-6314, USA

[‡] Equal contribution

[†] Corresponding author: ulf.peschel@uni-jena.de

# I) Structural Design and Fabrication

A sketch of the metal-insulator-metal (MIM) ENZ photonic waveguide, together with the optical excitation and propagation path, is shown in Figure S1a. The Gaussian beam is focused onto the in-coupling antenna that is connected to a 2D photonic ENZ waveguide (Fig. S1b, S2). Light propagates in TE-polarization (photonic mode, electric field in plane, Fig. S1a,c) through the waveguide and couples out at the out-coupling antenna. A scanning electron micrograph (SEM) of the fabricated structure is shown in Figure S1b. Figure S1c shows a cross-section of the MIM stack on the silica substrate that was created with a focused ion beam.

**Basic MIM waveguide**

We use polished silica quartz glass with a thickness of 170 $\mu$m as a substrate. The MIM waveguide layers were fabricated with a first PVD deposition (magnetron sputtering with AJA ATC-Orion-Series) of four layers in total (50 nm Cr, 1.5 $\mu$m Ag, 440 nm and 535 nm pure $SiO_2$, 1.0 $\mu$m Ag).

**In-coupling antenna for excitation**

The in-coupling-antenna configuration was then created with focused ion beam (FIB) milling (Zeiss-NVision 40 crossbeam) (Fig. S2a) to get an almost rectangular cross section of the dielectric waveguide channel (in-coupling antenna) of 420 nm or 460 nm in width and 440 nm or 535 nm in height. A second PVD deposition of two layers (800 nm Ag, 50 nm Cr) creates the metal sidewall at the in-coupling antenna (Fig. S2b), followed by a necessary sidewall FIB milling, which results in a 1D rectangular closed MIM waveguide, with a metal wall thickness of ca. 200 nm for the left and the right side (Fig. S2b). Finally, the in-coupling antenna was structured again with FIB in order to have an oblique excitation antenna configuration (Figs. S1a, S2c) with an angle $\phi = 54°$, with respect to the surface normal. The angle was chosen such that a perpendicularly incident focused laser beam is allowed to enter the dielectric channel.

**Out-coupling antenna for probing**

At the end of the MIM waveguide the out-coupling antenna was structured using FIB with the same cutting angle as used for the in-coupling antenna (Fig. S1b). This long oblique slit through all layers of the MIM waveguide was designed to scatter out the light along its extension, such that it can be clearly imaged in the far field with the experimental setup.

**Young's double slit, single and double hole**

Young's double slit was realized with three different pitches (p = 4, 6, 8 $\mu$m) by means of FIB cutting into the two dimensional waveguide (Fig. S3a, S4c-e). Also a single hole and a double hole as scattering objects were structured with the same procedure (Fig. S3b, S4a-b). To ensure a good signal to noise ratio at the CCD camera, a set of lines were scratched into the top layer to scatter away residual low-amplitude surface plasmon polaritons (SPPs) (Figs. S1b, S3a,b ) that otherwise would travel on the top surface of the top metal layer across the sample and scatter out at the out-coupling antenna. All performance-limited FIB cutting edges follow a maximum angle of $\phi \approx 6°$ (see Fig. S5), which for our experiments is sufficiently good.

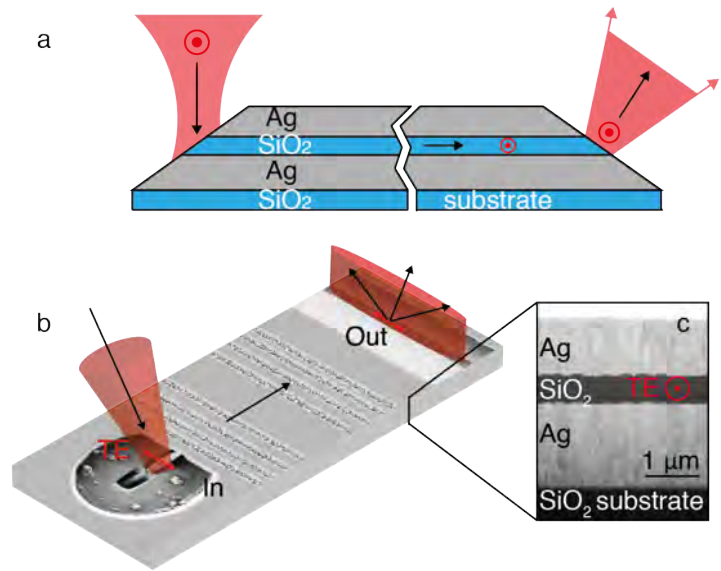

**Fig. S1: The 2D nanophotonic waveguide system**. **a)** Drawn sagittal cross-section of the system: A focused laser beam (red, left) with linear electric polarization is focused on a laterally confined resonant optical antenna. The antenna feeds a 2D photonic MIM waveguide. After propagating in the 2D waveguide light couples out at the out-coupling antenna, which is a laterally extended oblique cut through the MIM waveguide. **b)** SEM image in artificial 3D view of the fabricated structure. The in- and the out-coupling antennas are visible as well as the direction of light propagation (black arrows) and polarization of the electric field (red arrows). **c)** Cross-section (FIB) through the MIM waveguide with the direction of polarization of the electric field indicated.

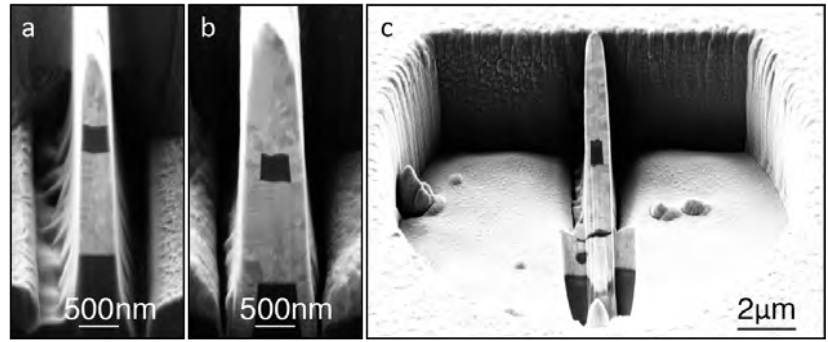

**Fig. S2: FIB milling process to manufacture the in-coupling antenna. a)** After the first PVD deposition of silver, glass and the top silver layer, the in-coupling antenna was structured using FIB milling. **b)** The second PVD-silver deposition process forms the silver sidewalls, which again need to be milled with the FIB in order to get straight walls. **c)** Finally the antenna is cut open with the FIB under an angle of 54° with respect to the perpendicular of the sample.

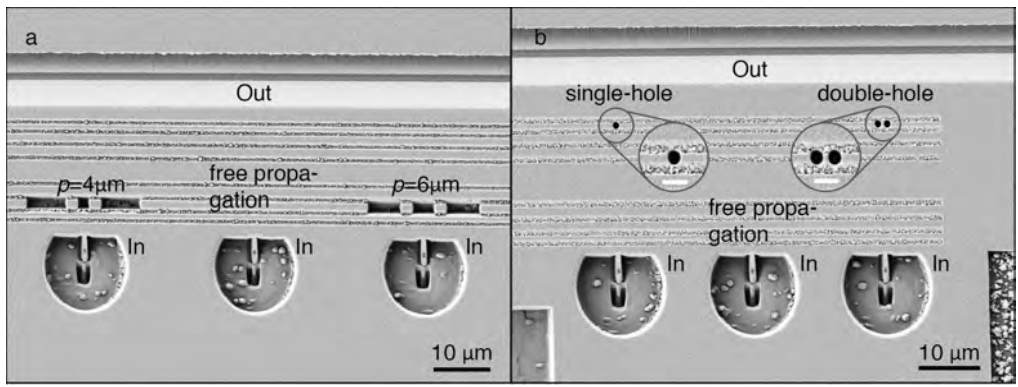

**Fig. S3: Scanning electron micrographs (SEM) of the fabricated structures in top-view. a)** Two double-slit structures inside the ENZ-MIM waveguide with two different pitches $p$ and a reference structure in the center (free propagation). **b)** Hole objects inside the ENZ-MIM waveguide. Left side: one hole (top (surface) diameter ø = 1 $\mu$m). Right side: two holes (top ø = 1 $\mu$m, pitch = 1.5 $\mu$m). Scale bar of the magnified insets: 2 $\mu$m.

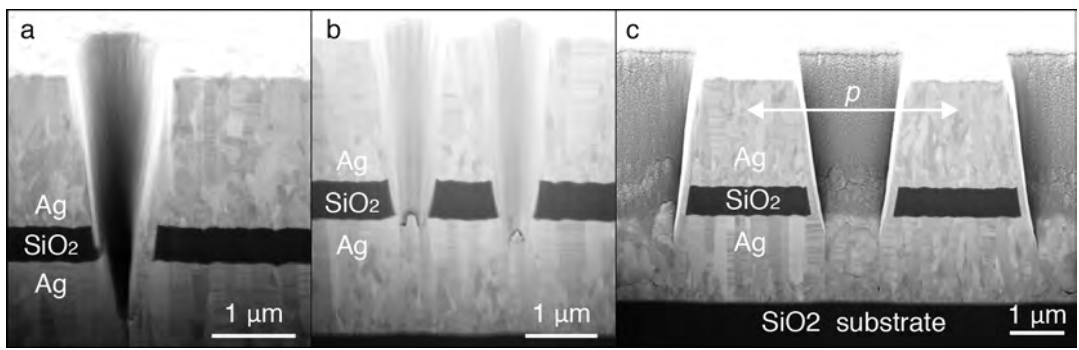

**Fig. S4: Meridional FIB cross-sections of the fabricated objects in the ENZ-MIM waveguide. a)** A single hole cut with a top-ø = 1.0 µm results in a conical hole with a lateral distance ranging from 365 nm towards 465 nm in the waveguide. **b)** Double hole cut with a top-ø = 1.0 µm for each hole and a top-pitch of *p* = 1.5 µm. The slightly conical shape of each hole leads to a slit size variation in the waveguide of 780 nm to 960 nm. **c)** Double slit with *p* = 4 µm.

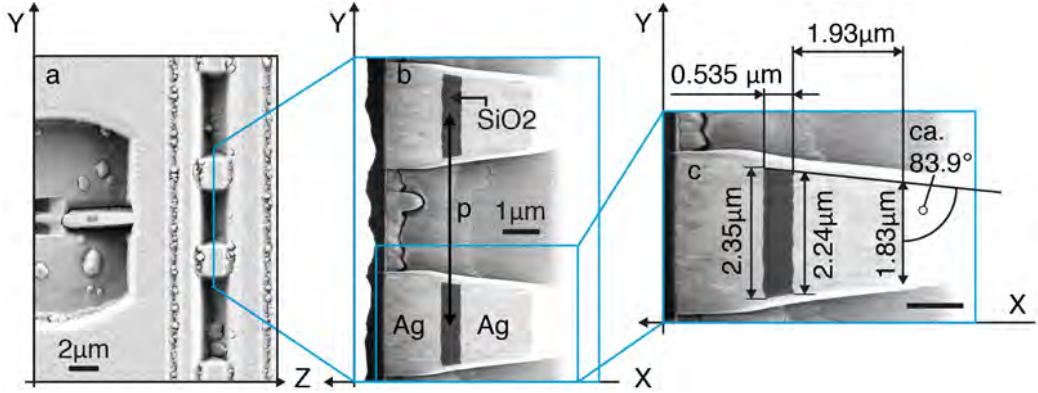

**Fig. S5: Meridional FIB cross section of the double slit. (a)** SEM image of the double slit. **(b)** Double slit cross section shows conical formed single slits with a pitch of p≈6 µm. **(c)** Single slit: Re-deposition during the FIB process leads to a cut angle of ϕ = 90°-83.9° = 6.1° (scale bar: 1 µm).

## Analytical calculation: Lowest order TE-mode cut-off for the MIM waveguide

The respective waveguide thickness $d^m_{cut\text{-}off,TE}$ for the m$^{th}$ order cut-off for a photonic TE mode (mode permittivity $\varepsilon_{eff\_TE}(\lambda_0)$=0) in a symmetric MIM waveguide can be calculated as [1]

$$d^m_{cut-off,TE}(\lambda_0) = \frac{\lambda_0}{\pi \cdot \sqrt{Re\{\varepsilon_D(\lambda_0)\}}} \cdot \left[\arctan\left[\sqrt{\frac{-Re\{\varepsilon_M(\lambda_0)\}}{Re\{\varepsilon_D(\lambda_0)\}}}\right] + m \cdot \pi\right],$$

with the vacuum wavelength $\lambda_0$ and the real term of the permittivity $\varepsilon_D$ and $\varepsilon_M$ for dielectric and metal, respectively. In our case, we chose the *m* = 0 order (lowest order). We use the mode dispersion in order to enforce the photonic TE-mode spectrally into its cut-off. Hence, the effective refractive index $n_{eff}(\lambda_0)$ reaches values close to zero. From this formula we calculate the curves for the waveguide core thickness that causes the cut-off of the lowest order mode (*m* = 0) over wavelength (Figure S6) in the

S5

wavelength range accessible with our experiment. We show the cut-off wavelength for the design thickness of the waveguide core ($d_0$ = 440 nm and $d_0$ = 535 nm) to be 1.4 $\mu$m and 1.67 $\mu$m, respectively. Also shown in the Figure is the experimental cut-off wavelength that we determined by calculating back from Young's experiment (see section IV) and the respective effective waveguide thickness $d_{Fit}$ that we determined by fitting. Both thicknesses, from theory and from the optical experiment deviate slightly, as we analyze in the article.

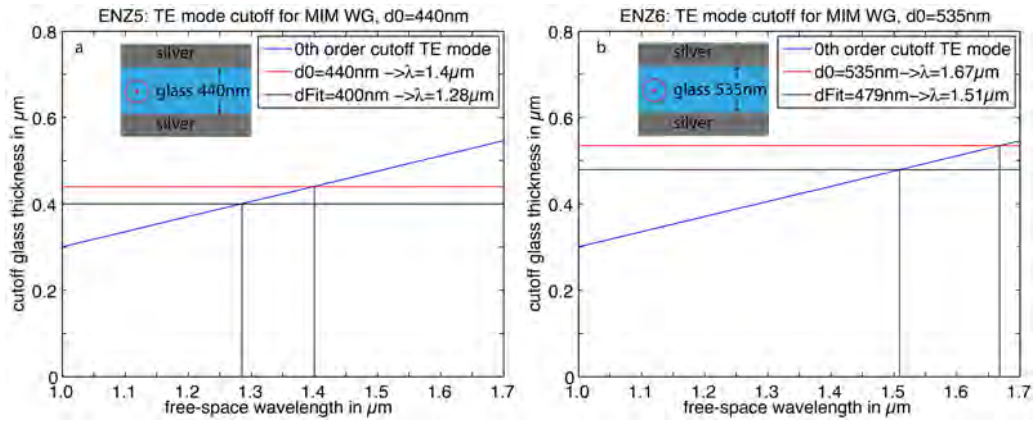

**Fig. S6: Analytical plot of the lowest order cut-off for the TE mode in the MIM waveguide with the respective analytical cut-off and the fitted cut-off wavelength. a)** The waveguide thickness $d_{SiO2}$ = 440nm and **b)** $d_{SiO2}$ = 535nm result in an analytical cut-off at $\lambda_0$ = 1.4 $\mu$m and at $\lambda_0$ = 1.67 $\mu$m, respectively. The analytical fit on the experimental result from Young's double slit gives an effective waveguide thickness of a) $d_{fit-SiO2}$ = 400 nm and b) $d_{fit-SiO2}$ = 479 nm for $\lambda_0$ = 1.28 $\mu$m and $\lambda_0$ = 1.51 $\mu$m, respectively.

# II) Optical measurement

The fabricated samples were placed on a 3D-piezo stage (manufactured by Physik Instrumente (PI) GmbH & Co. KG) in a custom-made optical setup (Fig. S7) that follows the far-field microscopic excitation and measurement principle published by Kriesch et al. [2] based on the experimental design published by Banzer et al. [3].

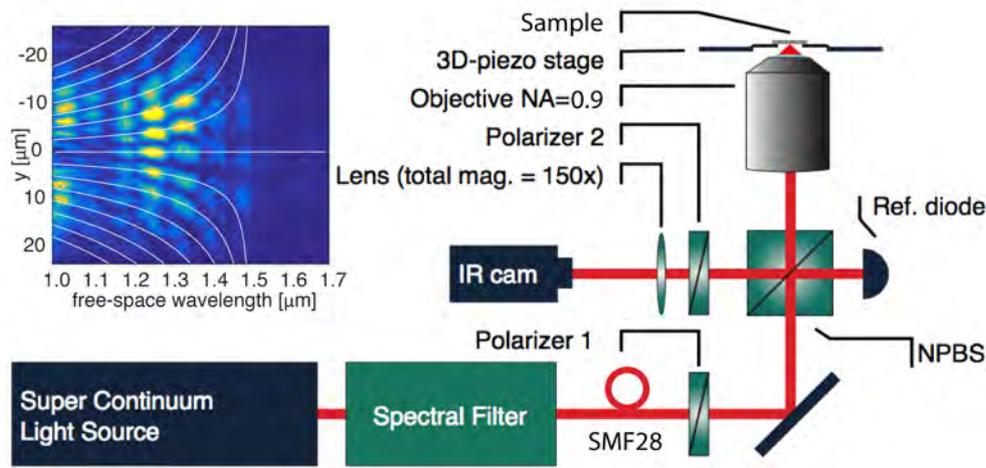

**Fig. S7: The experimental setup to probe the epsilon-near-zero (ENZ) photonic waveguides** for a NIR spectral range of $\lambda_0 = 1.0\ \mu m$ to $1.7\ \mu m$. The emission from the out-coupling antenna was imaged through polarizer 2 onto the InGaAs CCD camera (IR cam). (The setup sketch follows the design of A. Kriesch in [2]). Inset: The emission image from the out-coupling antenna is acquired successively while scanning over the wavelength range. Each image is automatically evaluated in the out-coupling antenna region and integrated to a lateral 1D intensity distribution (y axis of the graph) that is here plotted over the wavelengths. The figure in this case shows the diffraction orders of the double slit ($p = 8\ \mu m$, $d_{SiO2} = 535$ nm) diverging while the waveguide approaches ENZ towards its cut-off wavelength.

The collimated beam from a super continuum light source (SuperK Extreme EXR-15, NKT Photonics, $\lambda_0 = 0.48\ \mu m - 2.4\ \mu m$) is spectrally filtered by a programmed acousto-optic spectral tunable filter (AOTF, operated at $\lambda_0 = 1.0\ \mu m - 1.7\ \mu m$, resolution accuracy of $\Delta\lambda = 5$ nm, fwhm) and is then coupled into a single mode fiber SMF28. This fiber filters a fundamental Gaussian mode shape for the referred spectral range of $\lambda_0 = 1.2\ \mu m - 1.7\ \mu m$ and a predominantly Gaussian mode shape that is sufficient for some of the measurements for $\lambda_0 = 1.0\ \mu m - 1.2\ \mu m$. Pure transmission measurements, with respect to the in-coupling power, can only be performed in the spectral range of $\lambda_0 = 1.2\ \mu m - 1.7\ \mu m$.

The light is coupled out of the fiber, collimated and enters the experimental setup through a linear polarization filter (polarizer 1) and a non-polarizing beam splitter (NPBS) that directs 50 % of the power to a reference diode (InGaAs). The main beam (50 % of in-coupling power) enters a high NA microscope objective (Leica, HCX PL Fluotar 100x/0.90) and is focused to a diffraction-limited spot ($\o < 2\,\mu$m) that is adjusted with the piezos to cover the in-coupling antenna of the ENZ structure. Using the vertical displacement of the 3D-piezo stage the focus is adjusted on the in-coupling antenna. Light, which is propagated through the waveguide couples out and is then collected with the same objective. A NPBS reflects this imaging beam out of the excitation beam path and towards a second linear polarization filter (polarizer 2). For cross-polarization measurements polarizer 2 is set perpendicular to polarizer 1 with a maximum suppression of 1:10000. After polarizer 2, light passes the tubus lens to form a real image on an InGaAs NIR CCD camera (Xenics XS, 320 x 256 pixels).

The in-coupling polarization can be changed with polarizer 1 for selectively exciting the photonic TE mode or the plasmonic TM mode in the MIM waveguide. The rectangular shape of the in-coupling antenna supports both modes. The out-coupled light can be filtered with polarizer 2 in order to selectively detect the parallel- or cross-polarized component of the electromagnetic field that is emitted out of the out-coupling antenna.

## III) Simulation

Three-dimensional numerical simulations in the time-domain (FDTD) are performed in order to understand and verify results from the optical measurements on the ENZ-structures. The commercial software package Lumerical 3D FDTD Solutions® is used in order to simulate specific sections at once of the fabricated structure. A more computationally efficient, but less flexible in-house-developed code for 3D FDTD written in Fortran is used with the aim to cover the large-scale wave propagation through the ENZ structure with identical dimensions as in the experiment.

**Aim of the simulation**

The aim of the simulation is to investigate the free propagation of the photonic TE-mode in the bare ENZ waveguide and in a next step the diffraction of the photonic mode when placing a single-hole, a double-hole and a double-slit inside the waveguide. Both simulations deliver all complex field com-ponents for *E* and *H* taken from respective two-dimensional monitors. A peculiarity of the waveguide structure is the incorporated roughness at the two boundaries between glass and metal as a result of the bottom-up sputtering process (PVD), transferring the lower silver-layer roughness throughout the subsequently placed glass layer (TE-waveguide) to the adjacent top silver layer. Hence, the roughness features, located at the two material boundaries of the waveguide, are correlated to each other.

**Commercial software package Lumerical 3D-FDTD Solutions®**

With the software package Lumerical 3D FDTD Solutions® the influence of the roughness inside the waveguide is investigated for a smaller structure dimension of width x length x height = 8 $\mu$m x 10 $\mu$m x 0.535 $\mu$m (see section VIII on simulation results). In comparison, real dimensions as probed in the experiment are 50 $\mu$m (width) x 33 $\mu$m (length) x 0.535 $\mu$m (height). The roughness was incorporated in the simulation with a correlation length of 0.250 $\mu$m within the transversal area (X,Y), as determined from SEM images of waveguide cross sections. The noise amplitude is set to $\sigma_{RMS}$ = 0.018 $\mu$m, referring to a Gaussian distribution for the amplitude figure of merit. The distribution was varied over the boundary area using a random number generator. For the material dispersion for silver data from Johnson and Christy [4] and for the dispersion of silica data from Palik [5] are taken. Lumerical 3D FDTD Solutions® performs an internal fit for the material permittivity over the spectrum in order to fulfill the Kramers-Kronig relation. The simulation was performed for a wavelength range of $\lambda_0$ = 1.0 $\mu$m – 1.7 $\mu$m with a spectral resolution of $\Delta\lambda_0$ = 0.005 $\mu$m (equals 141 frequency points). The TE-mode is started inside the waveguide using a mode source over the complete width of the simulation domain. In the propagation direction as well as for the top and bottom, PML's covered the domain boundaries. In transversal direction, periodic-boundaries were added at both sides. The transmission results were exponentially extrapolated from the simulated length of $L$ = 10 $\mu$m to the experimental length of $L$ = 33 $\mu$m.

**In-house-written 3D-FDTD code to simulate the complete experimental structure**

The in-house-written 3D-FDTD code in Fortran on a large computing cluster allows for the simulation of the complete experimental structure at once. Particularly the crucial, but by size of the simulation domain challenging, diffraction simulations for single-holes, double-holes and for Young's double-slit in the ENZ domain are obtained in this way, approving and extending the Lumerical simulation results. In all simulations sufficient meshing and discretization settings were ensured and tested.

## IV) Extraction of $n_{\text{eff, TE}}(\lambda_0)$ from Young's double slit diffraction

Double-slit diffraction causes a periodic interference pattern. Leveraging this rule, for the evaluation of the diffraction pattern we utilize an evaluation algorithm that is based on the pattern frequency to improve signal to noise ratio, reduce systematic sample-induced errors and possible shifts.

**Spatial Fourier transformation**

For each measurement result from the double-slit experiments (Fig. 2, paper) and for each measured wavelength a Fourier transformation (i.e., for the measured spectral range of $\lambda_0 = 1.0\,\mu$m – $1.7\,\mu$m in steps of $\Delta\lambda_0 = 5$ nm a Fourier transform is performed for each of the resulting 141 wavelength points = (700nm/5nm)+1) of the intensities over the lateral coordinate $y$ is performed in order to translate the transverse distances (i.e. in $y$-direction) of the diffraction-maxima into frequency-space [$2\pi/y$]. The resulting frequency versus the free-space wavelength will be later on inserted into the intensity distribution [6] for the double-slit diffraction (equation (1)) in order to calculate the effective refractive index $n_{\text{eff}}$ out of the effective wavelength $\lambda_{\text{eff}}$.

$$I(y) = \frac{I_0}{2} \cdot \frac{\sin^2\left(\frac{\pi \cdot D}{\lambda_{eff} \cdot L} \cdot y\right)}{\left(\frac{\pi \cdot D}{\lambda_{eff} \cdot L} \cdot y\right)^2} \cdot \left(1 + \cos\left(\frac{2\pi \cdot p}{\lambda_{eff} \cdot L} \cdot y\right)\right) \qquad (1)$$

This intensity distribution in real-space [6] is a superposition of two functions with $\lambda_{\text{eff}}$ being the effective wavelength inside the ENZ waveguide and $L$ being the distance between the double-slit and the out-coupling antenna. The cosine function represents the diffraction orders with respect to the double-slit pitch $p$. The Sinc-envelope over all diffraction orders represents the diffraction at the single slit with the slit-width $D$.

Carrying out the Fourier transform is helpful to collect the maxima position of *all* lateral diffraction orders of the double-slit pattern (in $y$-direction versus free-space wavelength) at once by using the spatial frequency space. Hence, it is more efficient this way as to perform the evaluation in real space only.

**Paraxial correction**

Before starting with the Fourier transformation in lateral direction (i.e. in *y*-direction) for each measured wavelength ($\Delta\lambda_0 = 5$ nm, $\lambda_0 = 1.0\,\mu\text{m} - 1.7\,\mu\text{m}$), we conduct a paraxial correction to the experimental results as equation (1) is only valid within the paraxial limit. However, the paraxial limit is not fulfilled due to the geometry of our experimental probe. In the experiment the maximum captured orders refer to a diffraction angle above $\phi = 40°$ exceeding the diffraction angle for the paraxial limit of ca. $\phi = 22°$ by almost the factor of two.

Therefore, the emerging paraxial error $\Delta_{\text{ParError}} = y - b(y)$ has to be corrected for each wavelength where *y* is the maximum lateral position of the diffraction maxima in real-space and *b(y)* is the respective arc-length (Fig. S8). Each maxima position *y* has to be compressed towards the correct arc-length *b(y)* using equation (2)

$$b(y) = \arctan\left(\frac{y}{L}\right) \cdot L \qquad (2)$$

with *L* being the distance between the double-slit and the out-coupling antenna.

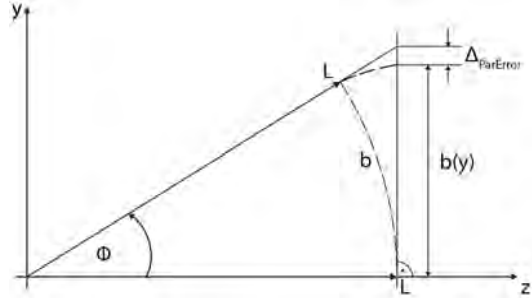

**Fig. S8:** Paraxial error $\Delta_{\text{ParError}}$, which has to be corrected.

After the paraxial correction and the subtraction of the background noise due to the measurement, we link the position of the 1st order maxima in k-space from the experiment with the respective cosine term in the intensity distribution for the double-slit (equation (1)). Hence, a Fourier transformation of the cosine term has to be performed before. For the following data evaluation it is easier to perform the transformation in frequency space ([f] = 1/y) and not in angular frequency space. The Fourier transform for the cosine term results in

$$f(t) = \cos(a \cdot t) \quad \rightarrow \quad \mathcal{F} \quad \rightarrow \quad F(f) = \frac{\delta\left(f - \frac{a}{2\pi}\right) + \delta\left(f + \frac{a}{2\pi}\right)}{2}$$

and the respective modulus frequency $f_0 = |f|$ of the delta function has to be

$$f_0 = \frac{a}{2\pi} \Leftrightarrow a = 2\pi \cdot f_0 \;.$$

The double-slit diffraction signal $f(t) = \cos(a \cdot t)$ in Fourier space together with a comparison of the cosine term from equation (1) leads to

$$f(t) = \cos(a \cdot t) \quad \Rightarrow \quad f(y) = \cos(2\pi \cdot f_0 \cdot y) \stackrel{\text{def}}{=} \cos\left(\frac{2\pi \cdot p}{\lambda_{eff} \cdot L} \cdot y\right).$$

Through mathematical comparison from the equation before, the frequency $f_0$ is

$$f_0 = \frac{p}{\lambda_{eff} \cdot L}$$

and subsequently the effective wavelength is

$$\lambda_{eff} = \frac{p}{f_0 \cdot L} \;.$$

The effective refractive index can now be calculated as

$$n_{eff}(\lambda_0) = \frac{\lambda_0}{\lambda_{eff}} \quad (3)$$

The complete and stepwise process from the measurement to the effective, refractive index can be seen in an example for one measurement for the thick layer and the pitch of $p = 8\,\mu m$ in Figure S9.

**The final $n_{eff}$ from Young's double slit experiment**

In general, the ensemble of all measurement results for $n_{eff}$, including the different double-slit pitches $p$ ($p = 4\,\mu m$, $6\,\mu m$, $8\,\mu m$) and two different waveguide thicknesses are taken in order to incorporate for the final $n_{eff}$. After having all $n_{eff}$ from each single experiment, a least square fit over all measurements per wavelength is realized. For the comparison with all measurements we use the analytical result for $n_{eff}$ with respect to the waveguide thickness $d_{eff}$ as the only fit parameter. Finally this results in the fitted refractive index $n_{eff}$ (fit) (see Fig. S9h). The detailed fit-result for the two different layer thicknesses is shown in table S1 together with a visual explanation in Figure S11 with respect to the incorporated roughness. The fitted thickness lies within the thickness range of the effective waveguide thickness, measured with the SEM (Fig. S11).

**Determination for $\lambda_{\text{cut-off}}$ in step-wise order**

The result can be seen, e.g., in Fig. S9h for $n_{eff}$(fit).

1. The double-slit experiment results in $n_{eff\,\text{experiment}}$ ($\lambda_1...\lambda_n$) for the measured vacuum wavelength from equation (3). For the thin/thick layer single measurements of the amount $i = 3/6$ were performed. Hence, for each wavelength $i = 3/6$ values for $n_{eff}$ are present and ready for the general fit with respect to the waveguide thickness $d_{eff}$.

2. An analytic function delivers the complex $n_{eff\,analytic}(d)$ for the photonic TE mode inside the unperturbed metal-insulator-metal waveguide.

    a. input: waveguide thickness $d$

    b. output: complex $n_{eff\,analytic}(d)_{\lambda 1\ldots\lambda n}$

3. The function Residual ($d$) is created, calculating the residual for $n_{eff}$ for the experimental results ("i" values per wavelength) for $n_{eff}$ versus the real part of the analytic results.

$$Residual(d)_{\lambda_1\ldots\lambda_n} = \left[ \sum_i \left[ n_{eff\,experiment,i} - Re\{n_{eff\,analytic}(d)\} \right]^2 \right]_{\lambda_1\ldots\lambda_n}$$

4. The function Residual($d$) will be minimized using the Matlab function "fminsearch". This results in the optimum thickness $d_{optimum}$ with respect to all measurement points $i$ per wavelength. The fit needs a start-value ($d_{start\text{-}value}$) for the thickness, which should be smaller than the optimum value. $d_{optimum} = fminsearch(Residual(d)_{\lambda_1\ldots\lambda_n}, d_{start-value})$

5. Taking $d_{optimum}$ and inserting it into point 2 results in the optimum $n_{eff}(\lambda_0, fit)$ showing spectrally the experimental cut-off (see Fig. S9h). For the thin/thick waveguide $d_{optimum}$ results in $0.400\mu m/0.479\mu m$ (see Tab. S1).

**Determination for $\Delta\lambda_{cut\text{-}off}$**

Having the experimental cut-off from Young's diffraction experiment the question is, what is the error of the fit (*fminsearch*) leading to $\Delta\lambda_{cut\text{-}off}$ This can be found in the evaluation of the chi-square-distribution in dependence of the thickness $d$ ($\chi^2(d)$). Here, we follow the achievments of Bevington and Robinson in [7] for setting up the $\chi^2(d)$ distribution and for finding its standard diviation resulting in $\Delta\lambda_{cut\text{-}off}$. The chi-squared-distribution, including the result for $d_{optimum}$, can be conducted to

$$\chi^2(d) = \frac{\sum_i \left[ n_{eff\,experiment,i} - Re\{n_{eff\,analytic}(d)\} \right]^2}{\sum_i \left[ n_{eff\,experiment,i} - Re\{n_{eff\,analytic}(d_{optimum})\} \right]^2}.$$

The result for the two layer thicknesses is shown in Figure S10. For the thin/thick layer the effective thickness results in $d_{eff}$ = 0.400 ± 0.012 $\mu$m / 0.479 ± 0.027$\mu$m. This leads to a cut-off wavelength including the error for $\lambda_{cut-off}$ = 1.29 ± 0.04 $\mu$m / 1.51 ± 0.08 $\mu$m.

**Determination for the analytical effective refractive index $n_{eff\_analytic}(\lambda_0)$**

As shown in section I, the respective waveguide thickness for the TE mode cut-off for the $m^{th}$ order can be calculated analytically under the requirement that the effective mode permittivity $\varepsilon_{eff\_TE}(\lambda_0)$ equals zero.

In the following, we need the general case, the respective waveguide thickness for the $0^{th}$ ($m$=0) order TE mode before the cut-off frequency (*1*), which can be conducted to

$$d_{TE} = \frac{\lambda_0}{\pi \cdot \sqrt{Re\{\varepsilon_D(\lambda_0)\} - Re\{\varepsilon_{eff_{TE}}(\lambda_0)\}}} \cdot \left[ \arctan\left[ \sqrt{\frac{Re\{\varepsilon_{eff_{TE}}(\lambda_0)\} - Re\{\varepsilon_M(\lambda_0)\}}{Re\{\varepsilon_D(\lambda_0)\} - Re\{\varepsilon_{eff_{TE}}(\lambda_0)\}}} \right] \right].$$

In order to get the analytical effective refractive index $n_{eff}(\lambda_0) = \sqrt{Re\{\varepsilon_{eff_{TE}}(\lambda_0)\}}$ one has to solve the following transcendental equation

$$\frac{d_{TE} \cdot \pi \cdot \sqrt{Re\{\varepsilon_D(\lambda_0)\} - Re\{\varepsilon_{eff_{TE}}(\lambda_0)\}}}{\lambda_0} - \arctan\left[ \sqrt{\frac{Re\{\varepsilon_{eff_{TE}}(\lambda_0)\} - Re\{\varepsilon_M(\lambda_0)\}}{Re\{\varepsilon_D(\lambda_0)\} - Re\{\varepsilon_{eff_{TE}}(\lambda_0)\}}} \right] = 0,$$

with the dependent variable $Re\{\varepsilon_{eff_{TE}}(\lambda_0)\}$, finding its root. As an input for the root calculation, the waveguide thickness $d_{TE}$ is constant with values being 0.440 $\mu$m/0.535 $\mu$m for the thin/thick waveguide. For trustworthy numerical results, the start values for $Re\{\varepsilon_{eff_{TE}}(\lambda_0)\}$ will be set on a straight line of the form

$$\varepsilon_{eff_{TE}}(\lambda_0) = m \cdot \lambda_0 + t.$$

The slope $m < 0$ is caluclated from two points, one from the experimental effective refractive index $n_{eff}(\lambda_0)$ result at $\lambda_0$ = 1.0 $\mu$m and the other from the analytical TE mode cut-off for $n_{eff}(\lambda_{cut-off})$=0 for both waveguide thicknesses, respectively. The two points are $P1$ = (1 $\mu$m/[$n_{eff}$(1 $\mu$m)]$^2$) and P2 = ($\lambda_{cut-off}$/0), respectively. The respective straight lines for the effective permittivity start values are

$$\varepsilon_{eff_{TE}}(\lambda_0) = -2.50 \cdot \lambda_0 + 3.50$$

for the thin wavguide layer (0.440 $\mu$m) and

$$\varepsilon_{eff_{TE}}(\lambda_0) = -1.81 \cdot \lambda_0 + 3.02$$

for the thick waveguide layer (0.535 $\mu$m). As a result a significant cut-off shift to shorter wavelengths for the experimentally determined effective refractive index in comparison to the analytical one can be seen in Figure S12. The reasons are discussed in the paper and origniating in general due to subwavelength perturbations between glass and metal, which will be discussed in more detail in section VIII (simulation results).

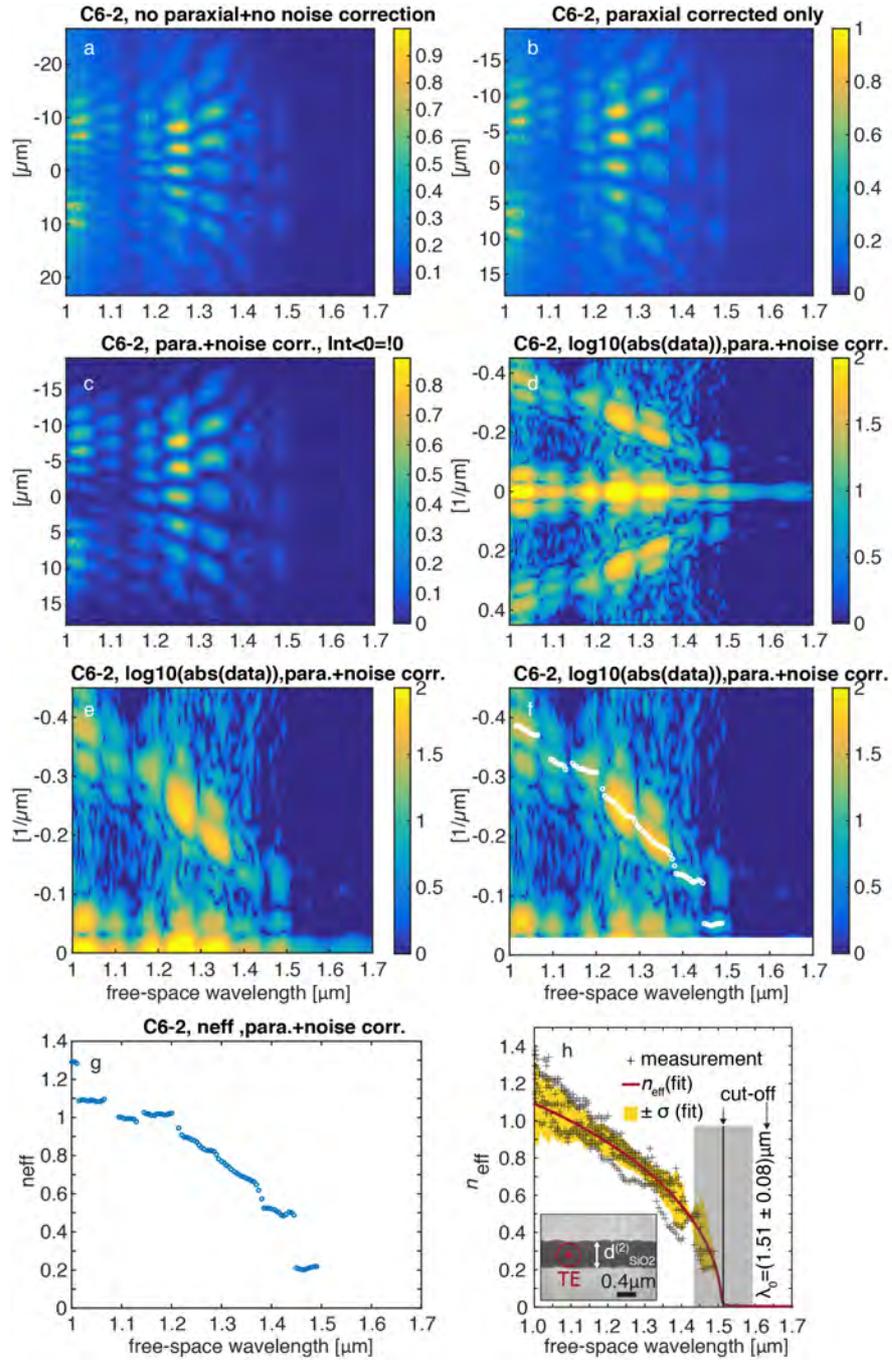

**Fig. S9: Stepwise extraction of $n_{eff}(\lambda_0)$ from Young's double-slit experiment.** It is shown for a double-slit with pitch $p = 8$ $\mu$m and d_SiO2 = 0.535 $\mu$m. **(a)** Diffraction result with neither noise correction nor paraxial correction applied. **(b)** Paraxial correction applied. **(c)** Paraxial correction and noise correction applied. **(d)** Wavelength wise (column wise) FFT from (c). **(e)** Plot of only one half space of (d). **(f)** Masking the $0^{th}$ order and find all maxima of the $1^{st}$ order (white dots). **(g)** Calculation of $n_{eff}(\lambda_0)$. **(h)** Result for $n_{eff}(\lambda_0)$ of six different double-slits leads after a least square fit with the waveguide thickness as the only fit-parameter to $n_{eff}$(fit).

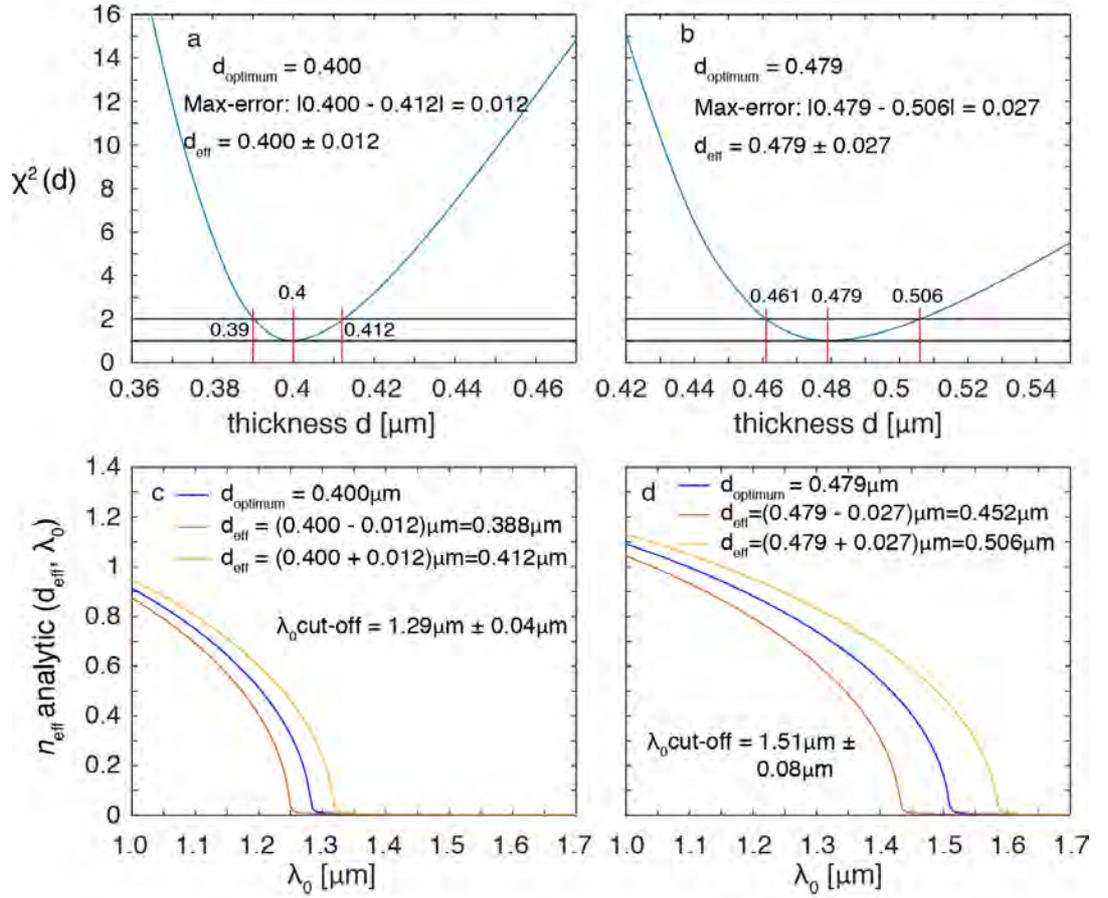

**Fig. S10: Determination of the Chi-square-distribution for the thin/thick ENZ waveguide in order to achieve the spectral error for the experimental cut-off** (a) Thin layer with a maximum thickness error of $\Delta d = \pm 0.012 \mu m$ (max difference for $[X^2(d) = 2 - X^2(d) = 1]$). This leads to a spectral error of (c) $\Delta \lambda_0 = \pm 0.04 \mu m$ around $\lambda_0 = 1.29 \mu m$. (b) Thick layer with a maximum thickness error of $\Delta d = \pm 0.027 \mu m$. This leads to a spectral error of (d) $\Delta \lambda_0 = \pm 0.08 \mu m$ around $\lambda_0 = 1.51 \mu m$.

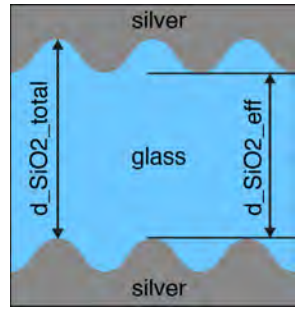

**Fig. S11: Idealized cross section of the ENZ waveguide structure**, comparable to the inset in Fig. S9h. This sketch shows a layer roughness in a non-realistic symmetrical manner in order to explain the findings in Tab. S1. The bottom and top metal layer are correlated to each other in terms of roughness amplitudes.

| waveguide layer | d_SiO2_total (SEM) | d_SiO2_eff (SEM) | d_SiO2 (fit) |
|---|---|---|---|
| thin | 440 nm ± 10 nm | 391 nm ± 10 nm | 400 nm ± 12 nm |
| thick | 535 nm ± 10 nm | 473 nm ± 10 nm | 479 nm ± 27 nm |

**Tab S1: Layer thicknesses for the two different manufactured layers.** The total thickness incorporates the perturbation between glass and metal, whereas the effective thickness excludes this roughness completely (see sketch in Fig. S11). The fitted thickness out of the double-slit diffraction is in good agreement with the effective thickness measured with the SEM.

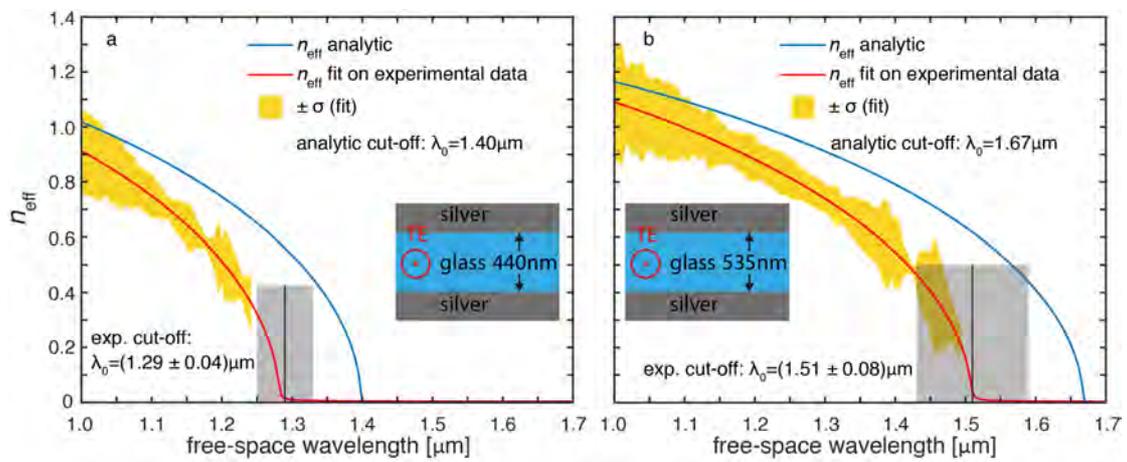

**Fig. S12: Analytic and experimental effective refractive index $n_{eff}$ for the photonic TE mode in the ENZ waveguide.** Cut-off shift between analytic and experimental is due to the perturbation inside the waveguide with the total thickness of **(a)** $d_1 = 0.440$ $\mu$m and **(b)** $d_2 = 0.535$ $\mu$m.

# V) Transmission measurements: single-hole, double-hole and double-slit

**Transmission measurement without objects inside the ENZ waveguide**

In the beginning the ENZ waveguide without objects is characterized by the excitation with either TE or TM polarized light (polarizer 1) as well as light emission from the out-coupling antenna is investigated in either TE or TM polarization (polarizer 2). The combination of all variations for the excitation- and detection polarization is shown in Figure S13 for two different structures for the thicker waveguide ($d = 535$ nm) and equal length. As a result, it shows that the in-coupling antenna blocks TM before TE polarized light and a second point is the conversion from TE into TM polarization, which happens inside the ENZ waveguide and not in the in-coupling antenna. These two points are important for the measurement, which will be discussed next.

**Diffraction measurements with objects inside the ENZ waveguide**

In Figure 3 of the main text the experimental results for wave propagation without an object, with a single-hole and a double-slit inside the waveguide are shown, which will be compared with simulations. In addition to that a detailed configuration of experimental- and simulation-results is shown in Figure S14 with respect to the cut-off shift into the blue. Here the unperturbed 3D-FDTD simulation result is spectrally fitted to the experimental result in order to find the new system cut-off.

As a result the cut-off shift for the single/double-hole is in the range of $\Delta\lambda_0 \approx 0.1\ \mu$m leading to a new cut-off of $\lambda_{0\_\text{cut-off}} \approx 1.67\ \mu$m $- 0.1\mu$m $= 1.57\ \mu$m (Figure S14 f,i). Whereas for the double-slit diffraction result the new cut off is $\lambda_{0\_\text{cut-off}} \approx 1.67\ \mu$m $- 0.12\mu$m $= 1.55\ \mu$m (Figure S14 l), the covering of experiment and simulation is more accurate due to a larger amount of diffraction orders.

**Transmission measurements with objects inside the ENZ waveguide**

In Figure S15 the experimental transmission results for objects in the ENZ waveguide are shown. Referring to the results shown in Figure S14, the final cut-off for the thick layer is at $\lambda_0 \approx 1.55\ \mu$m (see Fig. S15 a,d), which lies within the experimental cut-off range. The width of the measured intensity for free propagation is significantly growing starting at $\lambda_0 \approx 1.50\ \mu$m with a maximum at $\lambda_0 \approx 1.60\ \mu$m as a sign of field spreading in the spectral cut-off range (see Fig. S15 b). Moreover, the double-slit as an object becomes almost invisible due to the increase of the transmission ratio with vs. without double-slit from 20 % to almost 70 %, starting at the new cut-off at $\lambda_0 \approx 1.55\ \mu$m (see Fig. S15 e).

In summary, the diffraction measurements for all manufactured double-slits are shown in Figure S16 and the experimental cut-off range as well as the analytic cut-off is added to all images.

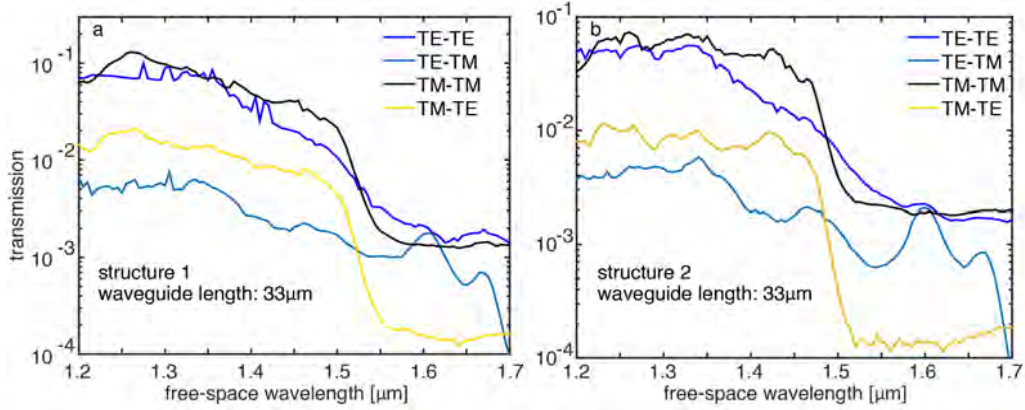

**Fig. S13: Experimental transmission result of an ENZ waveguide without objects and a length of $L = 33$ $\mu$m to characterize the in-coupling antenna.** Shown for two different structures: **(a)** structure 1 and **(b)** structure 2. "TE-TE" means: TE polarized excited at the in-coupling antenna (using polarizer 1) and TE polarized measured at the out-coupling antenna (using polarizer 2), etc. for "TE-TM" and so on. The cut-off of the in-coupling antenna is blue shifted for TM excitation compared to TE excitation. For "TE-TM" a transmission increase around $\lambda_0 = 1.60$ $\mu$m can be seen, which cannot be caused by the in-coupling antenna but happens inside the ENZ waveguide.

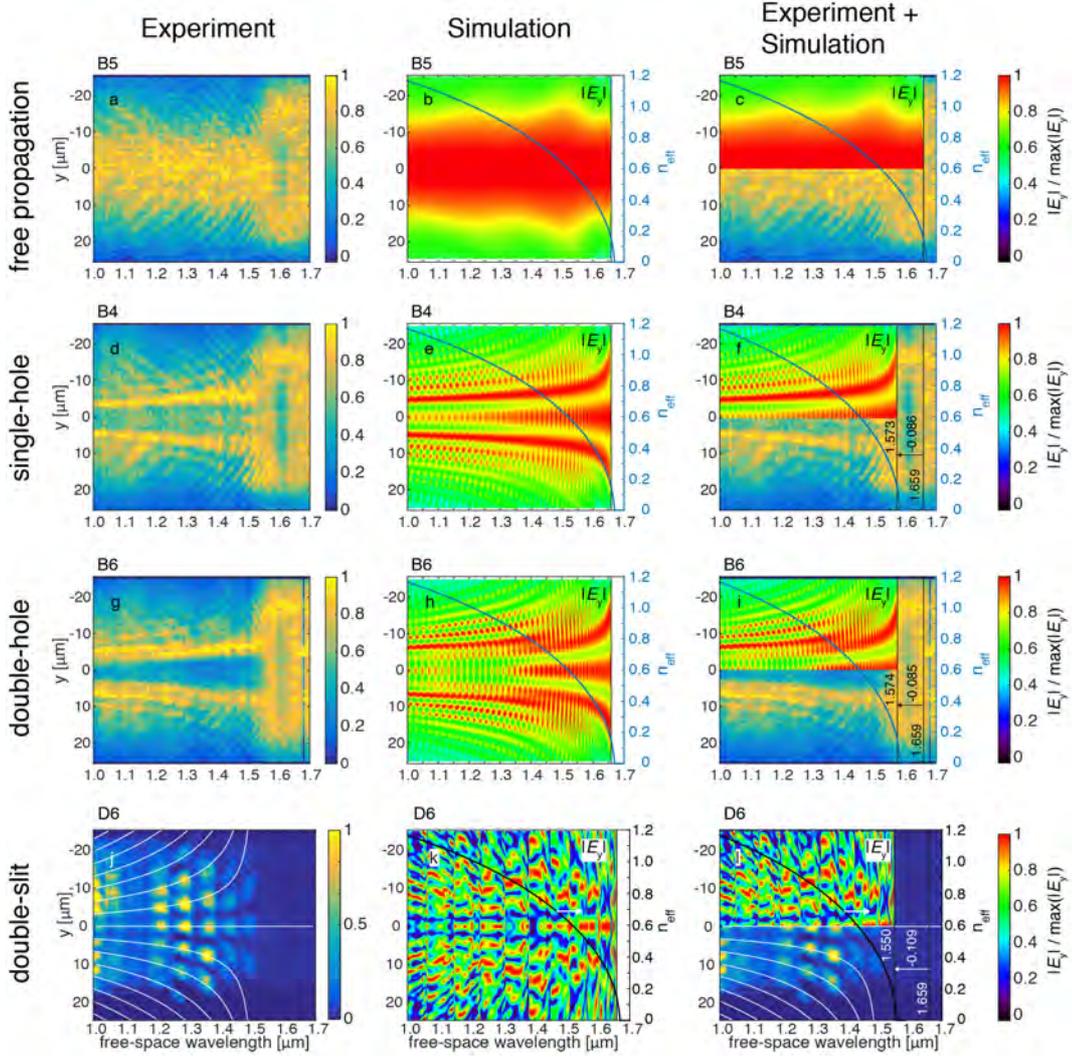

**Fig. S14: Objects in the thick ENZ waveguide in comparison with results from unperturbed 3D-FDTD simulations** ($d_{SiO2}$ = 535 nm). Waveguide length without objects: $L = 33$ µm. *Experiment*: normalized to the maximum per wavelength; TE-excitation and TE measurement. *Simulation*: spectral range from $\lambda_0 = 1.0$ µm to $\lambda_0 = 1.659$ µm. Analytical cut-off: $\lambda_0 = 1.67$ µm. The analytical effective refractive index $n_{eff}$ is added to the "Simulation" and "Experiment+Simulation" column. **(a-c)** Significant field narrowing in the simulation before the cut-off (b) and no cut-off shift offset tunable due to no presence of diffraction orders (c). **(d-f)** Strong 1st diffraction order present in both, experiment and simulation, as well as detectable higher orders. The single hole is placed 21 µm behind the in-coupling antenna (d,e). Cut-off shift with respect to the analytical cut-off: $\Delta\lambda_0 \approx -0.097$ µm. New cut-off: $\lambda_0\_cut\text{-}off \approx 1.67$ µm $- 0.097$ µm $\approx 1.57$ µm (f). **(g-i)** Stronger shadow effect due to a lack of the 0th order in the experiment compared to the single-hole (g). Diffraction orders are in good accordance between experiment and simulation (g,h). The double-hole is placed 21 µm behind the in-coupling antenna (g,h). Cut-off shift with respect to the analytical cut-off: $\Delta\lambda_0 \approx -0.096$ µm. New cut-off: $\lambda_0\_cut\text{-}off \approx 1.67$ µm $- 0.096$ µm $\approx 1.57$ µm (i). **(j-l)** Experiment: noise corrected image, not normalized to the maximum per wave-

length with pitch $p = 8$ $\mu$m (j). Double slit is placed 5 $\mu$m behind the in-coupling antenna (j,k). Cut-off shift with respect to the analytical cut-off: $\Delta\lambda_0 \approx -0.12$ $\mu$m. New cut-off: $\lambda_0\_cut\text{-}off \approx 1.67$ $\mu$m $-$ 0.12 $\mu$m = 1.55 $\mu$m (l).

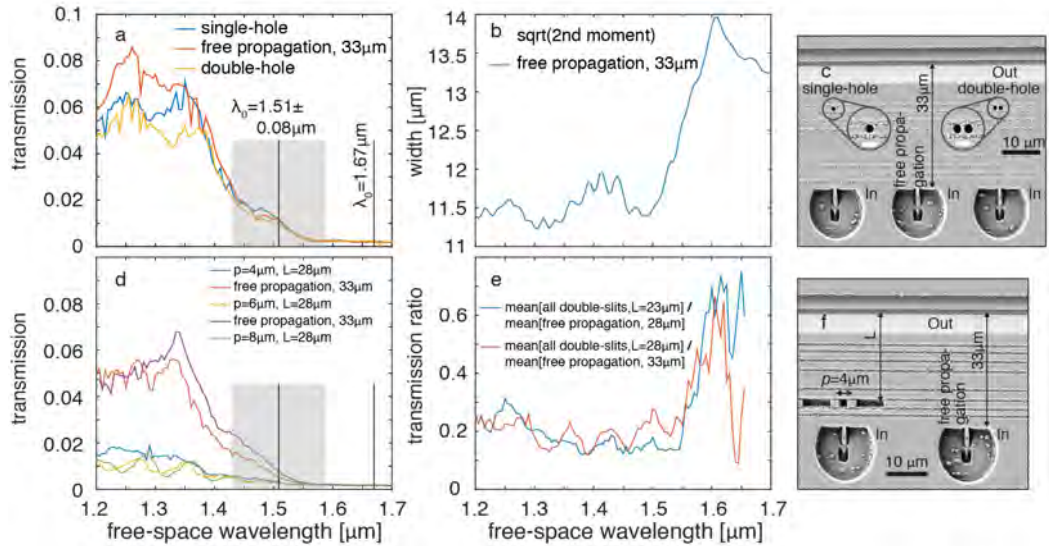

**Fig. S15: Experimental transmission results for objects in the thick ENZ waveguide ($d_{SiO2}$ = 535 nm).** Waveguide length without objects: $L = 33$ $\mu$m. **(a)** A single- and a double-hole in the ENZ waveguide. The respective cut-off lies within the experimental cut-off range, with a significant offset to the analytical cut-off at $\lambda_0 = 1.67$ $\mu$m. **(b)** The square root of the 2$^{nd}$ moment gives the width of the measured intensity for the free propagation with no object in the waveguide. The width rises significantly starting at $\lambda_0 = 1.51$ $\mu$m. **(c)** SEM image of the structures in (a-b). **(d)** Three different double-slits (see Figs. S16 (c,f,i)) in the waveguide and two structures without an object are shown. The significant difference in transmission with and without the double-slit is shown. **(e)** The transmission ratio of double-slits vs. no double-slits in the waveguide is shown. Transmission ratio results of Figures S16 (b,e,h) (blue) and of Figures S16 (c,f,i) (red) are shown. Both show an increase from ca. 20 % to ca. 70 % as a sign of vanishing objects (double-slit) in the waveguide.

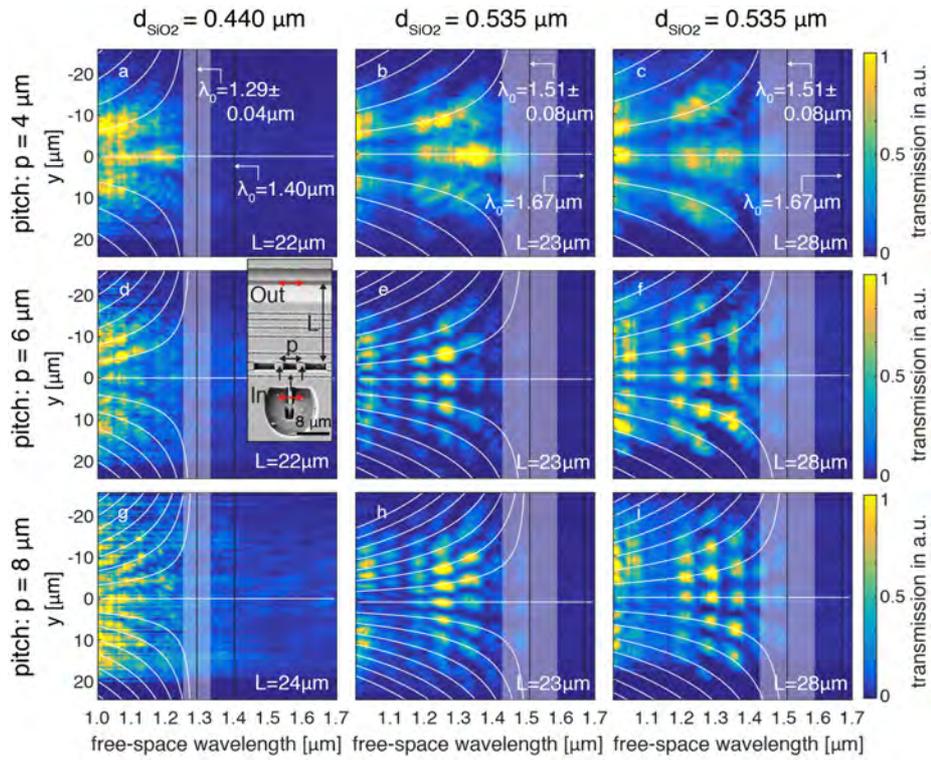

**Fig. S16: Result of Young's experiment in the thin/thick ENZ waveguide.** In all images the respective experimental and analytical cut-off ($\lambda_0 = 1.40\,\mu$m, $\lambda_0 = 1.67\,\mu$m) is added. Measurement: TE-excitation and TE-detection. Three double slit diffraction result shown for the thin waveguide **(a,d,g)**, for the thick waveguide with $L = 23\,\mu$m **(b,e,h)** and for the thick waveguide with $L = 28\,\mu$m **(c,f,i)**.

## VI) Noise Theory

To investigate the noise over the measured spectrum the experimental results are normalized for each measured wavelength, with respect to the sum of the intensity, which we call normalized measurement result or the norm. Let *u* be the measured signal and *du* the overlaid noise, the measurement represents

$$|u + du|^2 = |u|^2 + udu^* + u^*du + |du|^2 \equiv norm. \tag{4}$$

Let *u* and *du* be real numbers, equation (4) can be simplified to

$$|u + du|^2 - |u|^2 = u(du + du) + |du|^2 = u \cdot 2du + |du|^2 \quad u, du \in \mathbb{R}$$

$$\Leftrightarrow \frac{|u + du|^2 - |u|^2}{2u} = du + \frac{|du|^2}{2u} \tag{5}$$

and the noise-amplitude *du* together with the noise-error *|du|²/2u* can be extracted. For solving equation (5) one can fit with high accuracy a Gaussian curve

$$f(x) = a \cdot e^{-\left(\frac{x-b}{c}\right)^2} = |u|^2$$

with the amplitude *a*, the lateral offset *b* and the Gaussian width *c* onto the norm-measurement so that *|u|²* equals the Gaussian fit curve and *u* its square root. The amplitude *du* of the noise in its actual form is distributed around the zero line. Hence, this distribution can be considered equivalent to a field. For further investigation of the noise statistic (see section VII) one is interested in the intensity distribution, which leads to equation [6]

$$\left[\frac{norm - gaussian}{2 \cdot \sqrt{gaussian}}\right]^2 = \left[\frac{|u + du|^2 - |u|^2}{2u}\right]^2 = du^2 + du\frac{|du|^2}{u} + \frac{|du|^4}{4 \cdot |u|^2}$$

$$\Leftrightarrow \left[\frac{|u+du|^2 - |u|^2}{2u}\right]^2 \approx du^2 + \frac{du^3}{u} \tag{6}$$

with the negligible term $|du|^4/4|u|^2$. In the approximation of equation (6) the term $du^2$ is the noise intensity and the term $du^3/u$ its error. In summary the goodness of fit for the Gaussian fit curve as well as the noise amplitude (equation (5)) and noise intensity (equation (6)) are shown as an example in Figure S17.

**Validity of the noise equation**

One has to consider that equation (6) is only valid when the intensity noise $du^2$ is much smaller compared to the values of the Gaussian fit curve $|u|^2$, which is mostly not fulfilled at the boundary area, where the Gaussian curve approaches almost zero value. Simplified, one can state that the rule

$$|\text{norm} - \text{gaussian}| \ll \text{gaussian}$$

has to be fulfilled. Or in direct relation to the noise intensity of equation (6) the relative error

$$\frac{du^3}{u \cdot du^2} = \frac{du}{u} \ll 1.$$

Only within this approximation the extraction of the noise shows lowest errors. To give a reasonable error value of smaller than 10 %, the following approximation will be used for all measurement evaluations regarding the noise intensity of equation (6) for later statistic evaluation that is

$$\frac{du}{u} < 0.1 \quad \Rightarrow \quad du < \frac{u}{10} = \frac{\sqrt{\text{gaussian}}}{10}.$$

To fulfill this, one cannot use the total out-coupling antenna length of ca. $l = 50\ \mu\text{m}$ (in $y$-direction) and has to restrict that length symmetrically around the center. After a carful investigation of the data, this length was restricted to $\pm\ 10\ \mu\text{m}$ (example: Fig. S18).

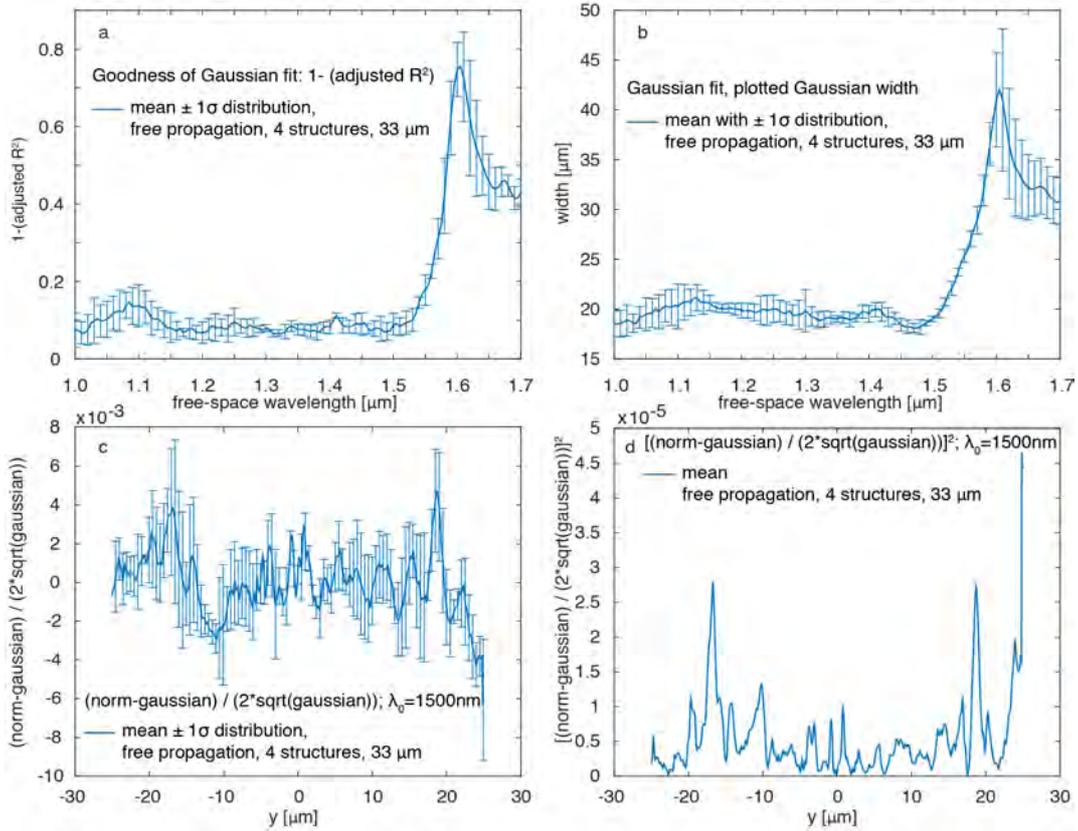

**Fig. S17: Gaussian fit onto the norm measurement of free propagation, noise amplitude and noise intensity.** **(a)** Goodness of fit (GOF, here zero value is the best-fit result) for the Gaussian fit onto the norm measurement (equation (4)) gets worse starting at $\lambda_0 = 1.55\,\mu$m, due to field spreading. **(b)** Fit result: Gaussian width. **(c)** Noise amplitude from equation (5) for $\lambda_0 = 1.50\,\mu$m. Trustworthy result for y = -10…0…+10 $\mu$m. **(d)** Noise intensity from equation (6) for $\lambda_0 = 1.50\,\mu$m.

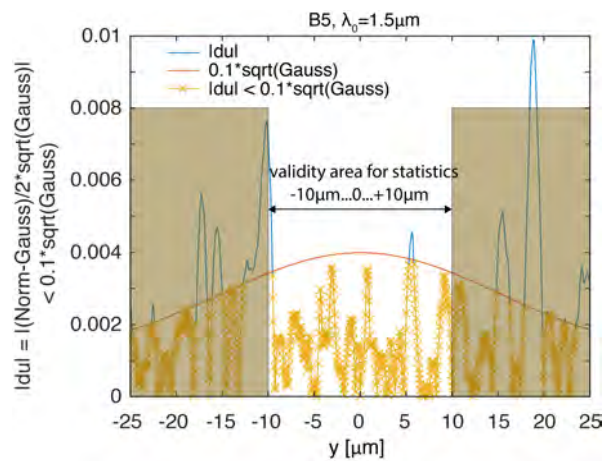

**Fig. S18: Restricted data evaluation for noise extraction with low error.** To achieve a small relative error of 10 % the out-coupling antenna length is restricted to -10 $\mu$m…0…+10 $\mu$m. Here an example for the wavelength $\lambda_0 = 1.5\,\mu$m is shown.

## Calculation of the noise strength: from experiment and from simulation

### Noise strength from experiment

Taking the intensity noise from equation (6) it is helpful to have a number representing the strength of the noise per wavelength. By summing up the results from different measurement over the wavelength and normalizing it to the norm measurement one results in the so-called noise strength per wavelength. For each of 4 different free propagating structures ($L = 33$ $\mu$m long) it is calculated to be

$$\left[ \frac{\sum_\lambda \left[ \frac{norm - gaussian}{2 \cdot \sqrt{gaussian}} \right]^2}{\sum_\lambda norm} \right]_i \quad with \ i = 4.$$

In the next step, the mean of the four results will be calculated to

$$noise\ strength = \overline{\left[ \frac{\sum_\lambda \left[ \frac{norm - gaussian}{2 \cdot \sqrt{gaussian}} \right]^2}{\sum_\lambda norm} \right]_i},$$

for i = 4, which is shown in the paper Figure 4a. There, the mean of the noise strength intensity is plotted together with the 1-σ-error distribution for every 2$^{nd}$ wavelength. A significant increase in the noise strength can be seen starting at λ=1.55$\mu$m.

### Noise strength from simulation: squared relative noise $\Delta f^2$

Six simulations with incorporated roughness in between the waveguide and the two metal sides were performed in Lumerical Solutions® (3D FDTD) as described in the simulation section III. Spectrally, the wavelength range from $\lambda_0 = 1.0$ $\mu$m to $\lambda_0 = 1.7$ $\mu$m in steps of $\Delta\lambda_0 = 0.005$ $\mu$m covered all simulations. From each simulation the squared relative noise $\Delta f^2$ is calculated and is averaged in a next step over all simulations. The result can be seen in Figure 4c of the main text in comparison to the measurement noise in Figure 4a.

In more detail, after a propagation length of 10 $\mu$m in the waveguide including the roughness the fields perpendicularly to the propagating z-direction were extracted calculating the Poynting vector component $P_z^{TE}$ for the photonic TE-mode to

$$P_z^{TE}(x, y, \lambda_0) = \frac{1}{2} \cdot Re\{-E_y(\lambda_0) \cdot H_x^*(\lambda_0)\}.$$

$\boldsymbol{P}_z^{TE}$ is a two-dimensional matrix for each wavelength, which will be reduced to a one-dimensional matrix over the wavelength by integrating the results over *x* to

$$P_z^{TE}1d(y, \lambda_0) = \int dx \; P_z^{TE}(x, y, \lambda_0).$$

We re-define the one-dimensional Poynting vector by the exchange of *y* by *n* to

$$f_n(\lambda_0) = P_z^{TE}1d(n, \lambda_0).$$

The squared relative noise *Δf²* can now be calculated for each wavelength by taking the difference of the averaged variance and the squared average with respect to $f_n$ versus the squared average of $f_n$ to

$$\Delta f^2(\lambda_0) = \frac{\langle \Delta f_n^2(\lambda_0) \rangle}{\langle f_n(\lambda_0) \rangle^2} = \frac{\langle f_n^2(\lambda_0) \rangle - \langle f_n(\lambda_0) \rangle^2}{\langle f_n(\lambda_0) \rangle^2} = \frac{\frac{1}{N}\sum_{n=1}^{N}\left(f_n(\lambda_0)\right)^2}{\left(\frac{1}{N}\sum_{n=1}^{N} f_n(\lambda_0)\right)^2} - 1$$

and its value can range from 0 at a minimum to 1 at the maximum indicating a 0 % speckle distribution towards a 100 % speckle distribution. As a result the squared relative noise is a vector for each simulation. At the end *Δf²* is averaged over all simulations, which is shown in the paper Figure 4c.

## VII) Noise statistics

Having the noise intensity from equation (6), one can calculate the normalized intensity using the average intensity $\langle I^2 \rangle$ as follows

$$\frac{I^2}{\langle I^2 \rangle} = \frac{\left[\frac{norm - gaussian}{2 \cdot \sqrt{gaussian}}\right]^2}{\left\langle \left[\frac{norm - gaussian}{2 \cdot \sqrt{gaussian}}\right]^2 \right\rangle}$$

to investigate, e.g., the distribution of higher intensity proportions over the width of the out-coupling antenna slit (see Fig. S19a-c). In a next step, the probability density function $p(x)$ with the constraint

$$\int_0^\infty p(x)dx = 1 \qquad (7)$$

is plotted in natural logarithm scale versus the normalized intensity (see Fig. S19d). One recognizes the attenuation of higher intensity portions $I^2$ with respect to the average intensity $\langle I^2 \rangle$. This attenuation follows a rule helping us to investigate whether this noise is a speckle noise or not.

**Proof of speckle distribution with the Chi-Square probability function**

For analysis, an exponentially decaying distribution cannot be taken to prove the presence of speckle because therefor always 2 independent and normally distributed parameters $X$ and $Y$ are necessary as an input. In our case that would be $Re\{du\}$ and $Im\{du\}$, which we do not have, as $du$ is a real number. Finishing that thought, one would get $Z = X^2 + Y^2$ and its exponential behavior follows the rule

$$p(Z') = \frac{|Z_0|}{Z_0} e^{-Z'} \quad \text{with} \quad Z' = \frac{Z}{Z_0}.$$

In the natural logarithm plot of the probability density function this turns out as a straight line. Our case is different as we only have one parameter $X$, which is $Re\{du\}$, we can consider for statistics. Hence, with the help of the Chi-Square probability function [8]

$$p_k(Z) = \frac{1}{2^{\frac{k}{2}} \cdot \Gamma\left(\frac{k}{2}\right)} \cdot Z^{\frac{k}{2}-1} \cdot e^{-\frac{Z}{2}} \quad \text{with} \quad Z = X^2 \text{ and } k = 1$$

it is proven that $X^2$ follows that function. Hence, $X$ is normally distributed and the noise is explicitly a speckle noise. To test this, we plot this distribution with $Z = X^2$ together with the noise probability density distribution and find a very good agreement not only for low but also for intensities, which are up to four times the average intensity $\langle I^2 \rangle$ (see Fig. S19d).

Deviations from the Chi-Square distribution occur for larger intensities indicating the error in statistics. The reason for that error is the small start ensemble of four structures (free propagation) leading to rare probability events for large intensities, which would increase in probability when the start ensemble would be larger. The rarest probability event, which can occur, is shown in Figure S19d by the lowest horizontal line. The ascending lines representing the stepwise increasing probability. Those kinds of error lines can be calculated analytically. To start with, one needs the normalization factor $N$, which is necessary to normalize the counts of the histogram per bin (stepsize of $I^2/\langle I^2 \rangle$ ) with the probability density function $p(x)$ to

$$p(x) = \frac{counts(x)}{N}$$

in order to fulfill equation (7). Before, the normalization factor $N$ has to be calculated to

$$N = \int counts(x) \cdot dx ,$$

where the counts per bin are multiplied with the bin-step size (dx) and summed up over all averaged intensities. The probability figure of merit $p_{min}$ for the horizontal line for the rarest probability event of four structures of same length is calculated to

$$p_{min} = \frac{1}{4 \cdot N}$$

The respective ascending probability lines have values of

$$u \cdot p_{min}$$

with $u = [2,3,4...]$ (Fig. S19d).

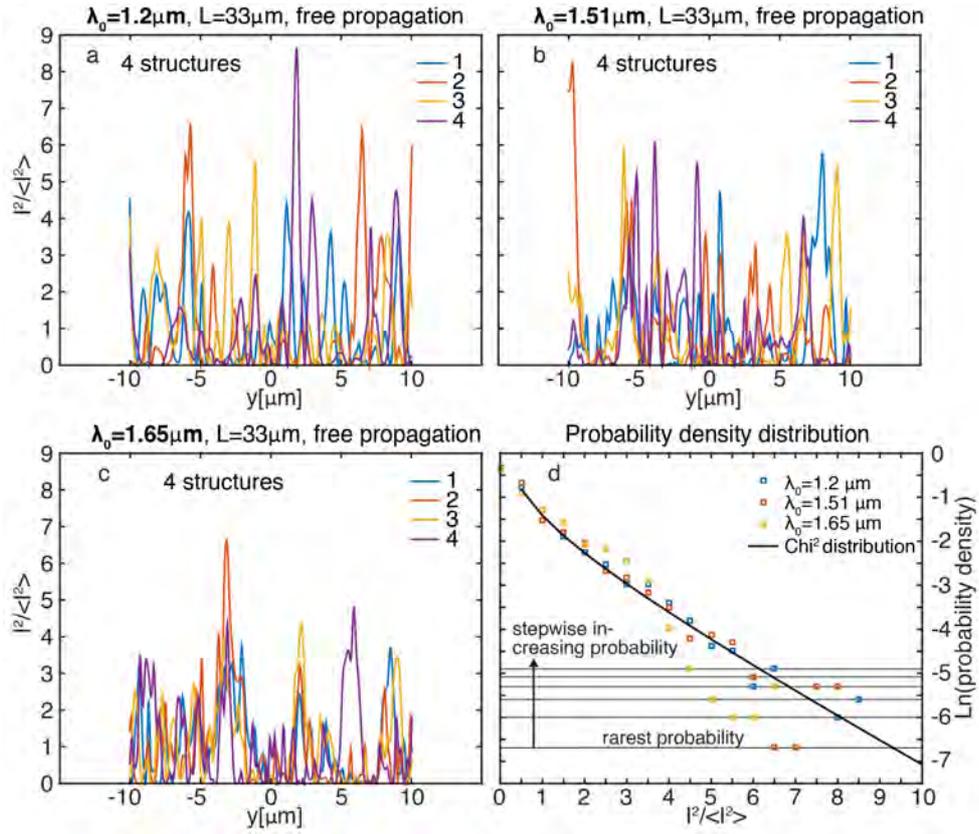

**Fig. S19: Normalized intensity and probability density distribution for the extracted noise.** **(a-c)** Normalized intensity for $\lambda_0 = [1.2\,\mu m, 1.51\,\mu m, 1.65\,\mu m]$ vs. the out-coupling length of the antenna in the restricted data range (see Fig. S16). **(d)** Probability density distribution (averaged over four structures) follows the Chi-Square distribution as a prove for speckle noise. Deviations from the Chi-Square distributions are due to the small start ensemble (4 structures) leading to rare and rarest probabilities.

## VIII) Simulation results

**FDTD simulations for the complete ENZ-waveguide dimension without perturbations**

As introduced in section III, the numerical simulations for the ENZ-structure with real dimensions were implemented with a self-written 3D-FDTD code in Fortran (domain sketch in Fig. S20). This also includes the dimension of the in-coupling waveguide towards the ENZ waveguide. As a simulation result the field distribution with and without objects being present inside the waveguide (hole, double hole, double slit vs. free propagation) will be discussed next.

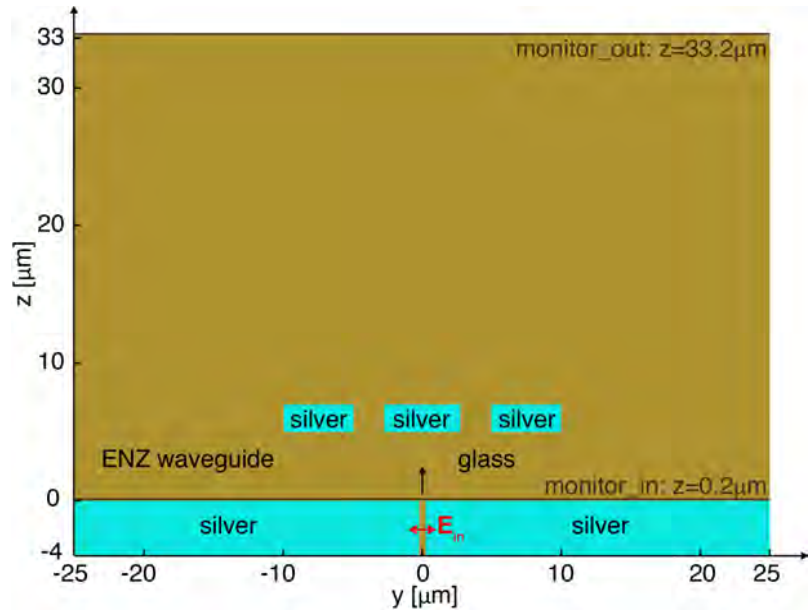

**Fig. S20: 3D FDTD domain for the ENZ waveguide in real dimensions for x=0.** Example for the double slit ($p=8\mu$m) and a waveguide length of 33$\mu$m. The respective transmission is calculated by the ratio of the respective power (Poynting vector in z-direction) of monitor_out vs. monitor_in. The in-coupling field is in TE polarization (y-direction).

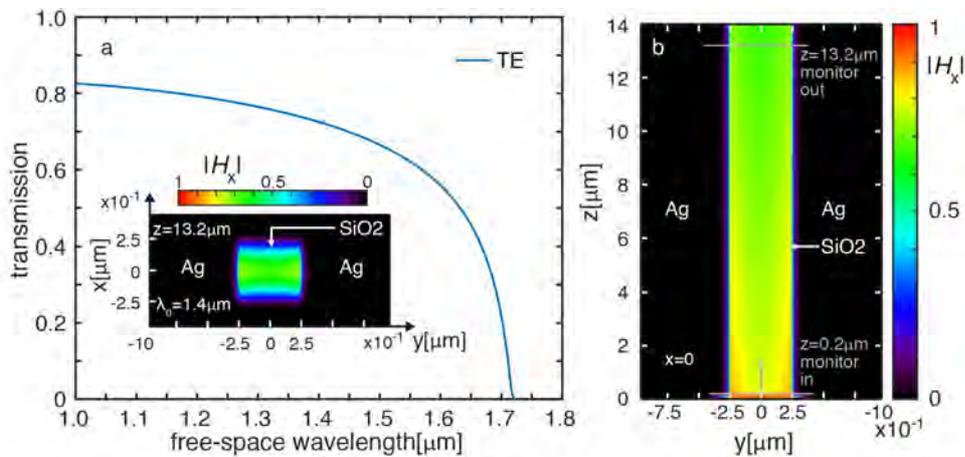

**Fig. S21: Polarization resolved transmission investigation over the spectrum for the in-coupling waveguide (3D FDTD).** Excitation of the TE mode in y-direction at z=0 at the waveguide with the length of 14 $\mu$m, pumping with the H-field. Transmission investigation for a length of 13 $\mu$m. Mode profile for $\lambda_0 = 1.4$ $\mu$m **(a)** TE cut-off at $\lambda_0 \approx 1.72$ $\mu$m and a TM polarization ratio of smaller than $3 \times 10^{-5}$ with respect to the total transmission. Stationary mode distribution in the xy-plane at z = 13.2 $\mu$m and in **(b)** yz-plane at x = 0.

**Simulation results for the in-coupling waveguide**

The in-coupling waveguide is investigated numerically by exciting the TE polarized plasmonic mode (y-direction) in order to find spectrally the TE cut-off and the ratio of polarization conversion into TM polarized light (Fig. S21). As a result, the TE cut-off is at $\lambda_0 \approx 1.72\,\mu$m (Fig. S21a) and the conversion ratio into TM is smaller than $3 \times 10^{-5}$ with respect to the total transmission. Hence, no TM polarized light will be launched into the ENZ waveguide with respect to the in-coupling waveguide.

**Simulation results for the complete ENZ-waveguide dimension – equal to the experimental probe**

The respective ENZ waveguide cut-off for TE ($\lambda_0 = 1.67\,\mu$m) is blue shifted (Fig. S24a,b: no object) in comparison to the in-coupling waveguide cut-off (Fig. S21a). Therefore, the ENZ waveguide leads to the cut-off of the photonic TE-mode rather than the in-coupling waveguide with its plasmonic TE-mode.

Figure S22 shows field distributions inside the ENZ waveguide for free propagation as well as diffraction at the implemented single- and double-hole. The in-coupling waveguide is shielded in those plots as well as some distance of the adjacent ENZ waveguide due to oversaturation. As can be seen in Figures S22a,d,g the emission into the ENZ waveguide shows a dipole like field distribution and a reduced field strength with increasing wavelength. Objects with the same dimension like in the experiment lead to diffraction pattern in the dipole field cone, which is shown for the single hole (Figs. S22b,e,h) and the double hole (Figs. S22c,f,i).

**Diffraction plots for the double-slit simulation in the ENZ waveguide**

Spectral plots for the double slit show the increasing amount of diffraction orders when the slit width is increased from $p = 6\,\mu$m to $p = 8\,\mu$m showing diffraction orders until the $\pm 4^{th}$ and the $\pm 5^{th}$ order, respectively (Fig. S23). Vertical lines show Fabry-Pérot resonances in between the slits and the backside metal before the double slit (see Fig. S20).

**Transmission results for objects in the ENZ waveguide**

Respective transmission spectra for objects in the waveguide show that little holes do not influence the total transmission with respect to no objects being present (Fig. S24a). The double-slit leads to a strong polarization conversion from TE to TM (Fig. S24b,c), compared to a small one for the holes as they almost coincide with the transmission line for free propagation (no objects, Fig. S24a). The reason for the polarization conversion of the double-slit is the large area of conical cuts (conus angle $\phi \approx 6°$) for the double slit geometry taken from the fabrication process (see Fig. S5). This conical angle was implemented in the 3D FDTD simulation.

**Poisson spot**

Also, the simulation reproduces the Poisson spot in a field line plot at the out-coupling antenna after a waveguide length of L = 33 $\mu$m (Figs. S25a-c). The double-hole acts as a connected shield bar for light as no light passes the slit in between the two holes. As a result, the contrast in diffraction at the out-coupling antenna is slightly larger (Figs. S25d-e). The diffraction pattern as well as the Poisson spot vanishes close to the analytical cut-off for the single-hole (Fig. S25f).

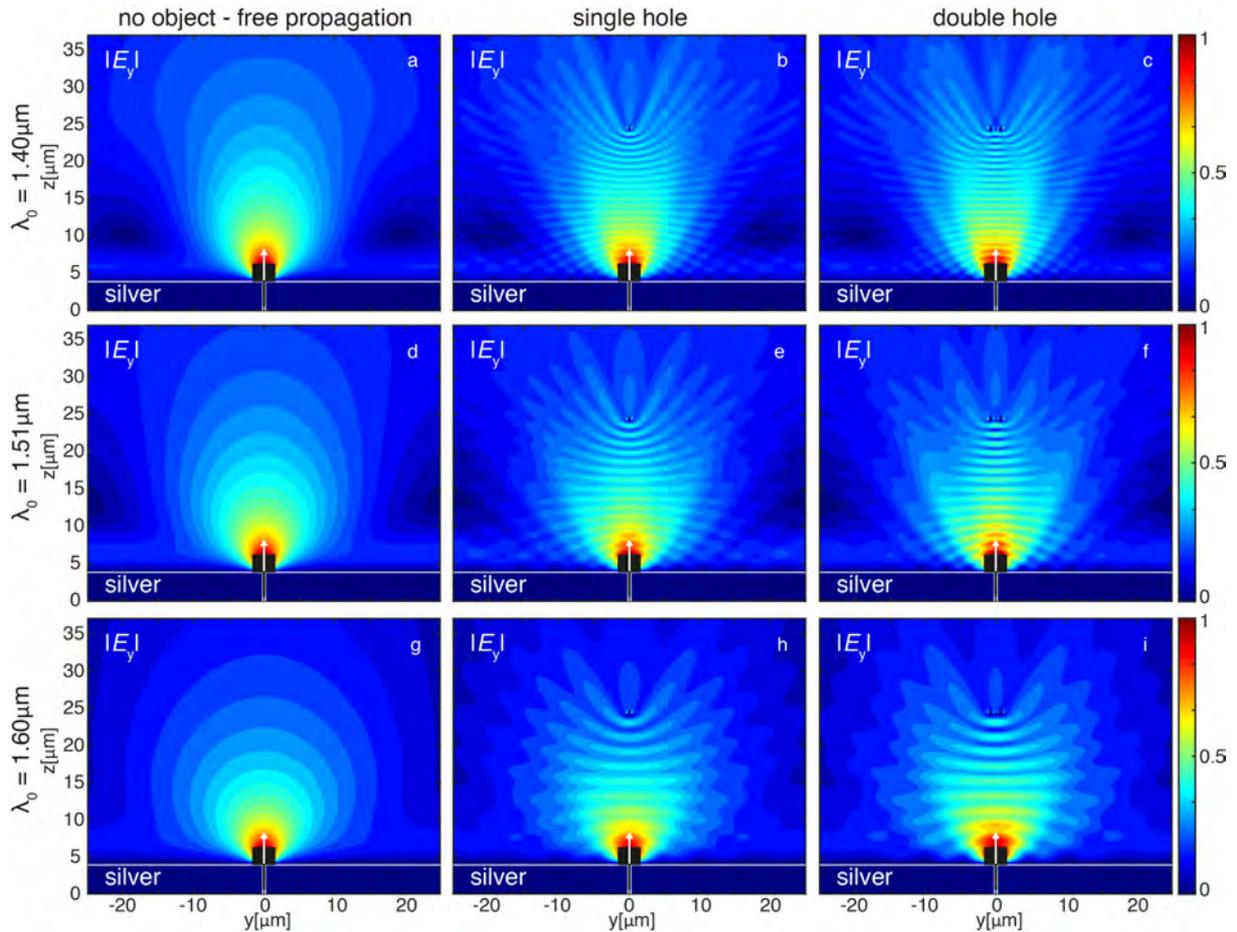

**Fig. S22: Simulated field profiles (modulus of the complex $E_y$ field) in 3D-FDTD.** Field values for the incoupling antenna (0.4 $\mu$m in y-direction and 4 $\mu$m in z-direction) as well as a small part of the waveguide are removed due to data over saturation. **(a,d,g)** Dipole like field distribution (dipole axis in y-direction) for the free

propagation and reduced field strength with growing wavelength. **(b,e,h)** Interference pattern due to diffraction at the single hole and the double hole **(c,f,i)** in the waveguide.

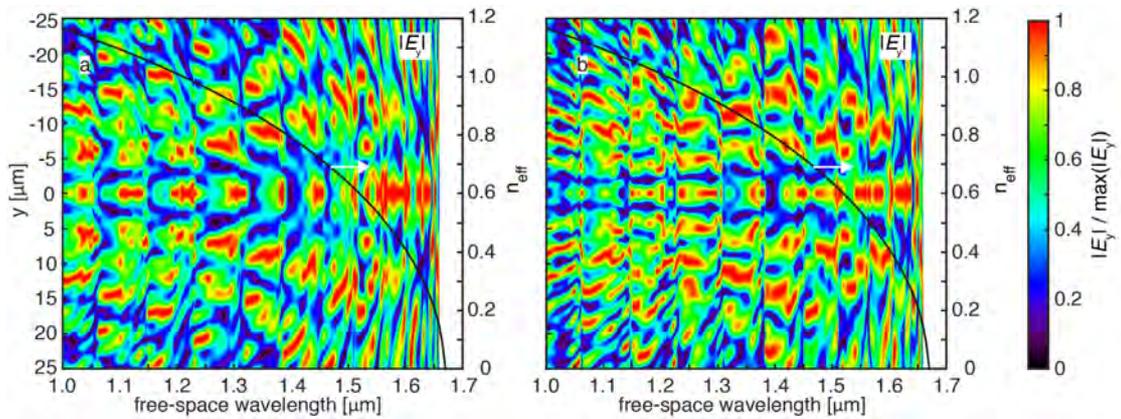

**Fig. S23: Spectral plots for the double slits inside the waveguide ($L = 33\mu$m) with the analytical effective refractive index $n_{eff}$.** Both plots are wavelength wise normalized to its maximum. **(a)** Pitch $p = 6$ $\mu$m, diffraction until the $\pm 4^{th}$ order. **(b)** Pitch $p = 8$ $\mu$m, diffraction until the $\pm 5^{th}$ order.

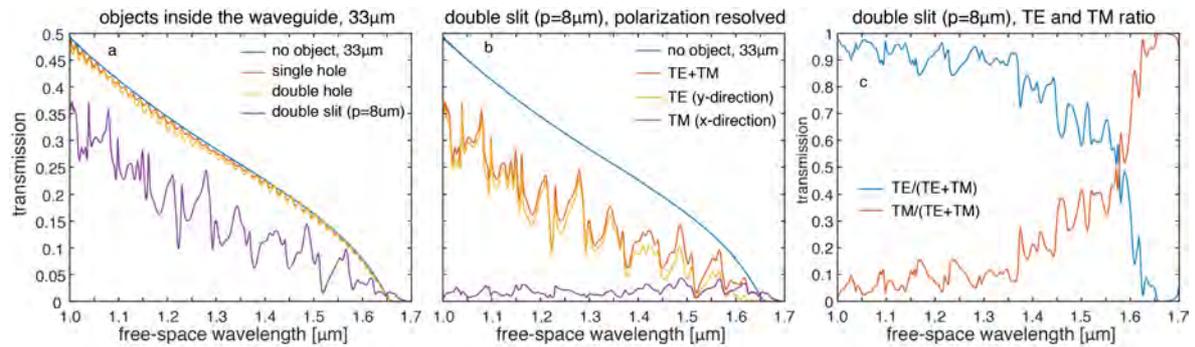

**Fig. S24: Transmission plots for objects in the ENZ waveguide. (a)** The hole and double-hole do not reduce the transmission compared to the case without object. The double-slit shows a significant transmission reduction and even a transmission beyond the analytical cut-off (1.67 $\mu$m) due to polarization conversion. **(b)** Polarization conversion from TE into TM for the double slit with **(c)** the plotted ratio with respect to the total transmission.

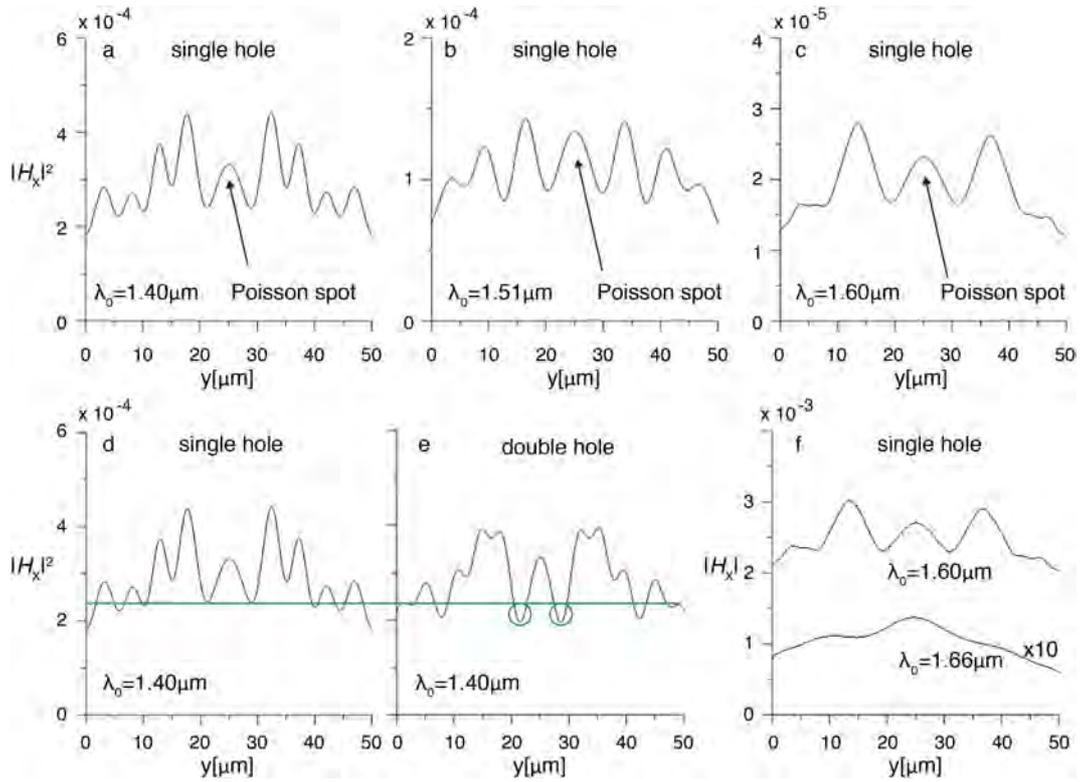

**Fig. S25: Poisson spot, contrast difference and vanishing hole object for $L=33\mu$m in the ENZ waveguide (d=535nm). Squared modulus of the complex $H$-field at the out-coupling antenna for light propagation in z-direction. (a-c)** The Poisson spot as the zeroth diffraction maximum is clearly visible. The shadow region beside the Poisson spot does not reach zero field value. **(d-e)** Contrast difference for $\lambda_0=1.40\mu$m between the single and the double-hole (pitch=$1.5\mu$m). **(f)** (modulus of the complex $H$-field) The hole object almost vanish close to the analytical cut-off of $\lambda_0=1.67\mu$m.

**3D FDTD with incorporated perturbations (FDTD Lumerical Solutions®)**

Three-dimensional (3D) numerical simulations with incorporated perturbations in between metal and glass of the MIM ENZ waveguide are performed with the commercial FDTD Lumerical Solutions® software package as described in section III. A sketch of the simulation domain in the XZ area for Y = 0 can be seen in Figure S26.

**Correlated perturbations, TE-mode**

The perturbations were implemented using a surface roughness script from Lumerical Solutions®. The respective correlation length was analytically calculated from a Fourier transform of the metal-glass boundary SEM image shown in Figs. S1c,S4. In a next step the amplitude in x-direction was chosen that the effective waveguide thickness fits the one measured with the SEM to $d = 473$ nm (see Fig. S11 and Tab. S1). The lateral distribution (in x and y) of the perturbation was varied using a random number generator. In x-direction the absolute thickness, defined as the distance between the two metal boundaries, always fits the waveguide thickness of $d_{\text{glass}} = 0.536\ \mu$m (Fig. S11 and Tab. S1). Six simulations with a different lateral perturbation distribution were simulated and averaged.

**Results**

Figure S27 shows the transmission results for each of the six simulations versus the experimental results for the thick waveguide ($d = 535$ nm). The simulation shows the mean transmission, exponentially extrapolated from the numerical result for the propagation length of $L=10\mu$m to the experimental distance $L=33\mu$m. The total transmission $T$ and the transmission in $TE$ (y-direction), both in propagation direction (z-direction), are calculated by integrating the respective Poynting vector over the XY-monitor area to

$$T = \frac{1}{2} \iint dx\, dy\, Re\{P_z\} = \frac{1}{2} \iint dx\, dy\, Re\{E_x \cdot H_y^* - E_y \cdot H_x^*\} \text{ and}$$

$$TE = \frac{1}{2} \iint dx\, dy\, Re\{P_z(E_y)\} = \frac{1}{2} \iint dx\, dy\, Re\{-E_y \cdot H_x^*\},$$

both normalized to the power of the mode source. The transmission in $TM$ cannot be exponentially extrapolated. Also, it is the only polarization direction in which $TE$ can be converted. Hence, the difference between $T$ and $TE$ can be calculated from the previous results to

$$TM = T - TE.$$

The cut-off shift as a result of the incorporated perturbations can be seen in *TE* polarization in Figure S27b. Therein the TE-transmission result for having no perturbation in the waveguide versus the mean TE-transmission result with perturbation is shown. When *TE* runs into cut-off a polarization conversion into *TM* polarization can be seen in both, experiment and simulation (see Figs. S27c-f). Hence, the perturbation leads to a cut-off shift into the blue. This spectral cut-off range from the simulation coincides with the experimentally obtained cut-off range of $\lambda_0 = 1.51\ \mu m \pm 0.08\ \mu m$.

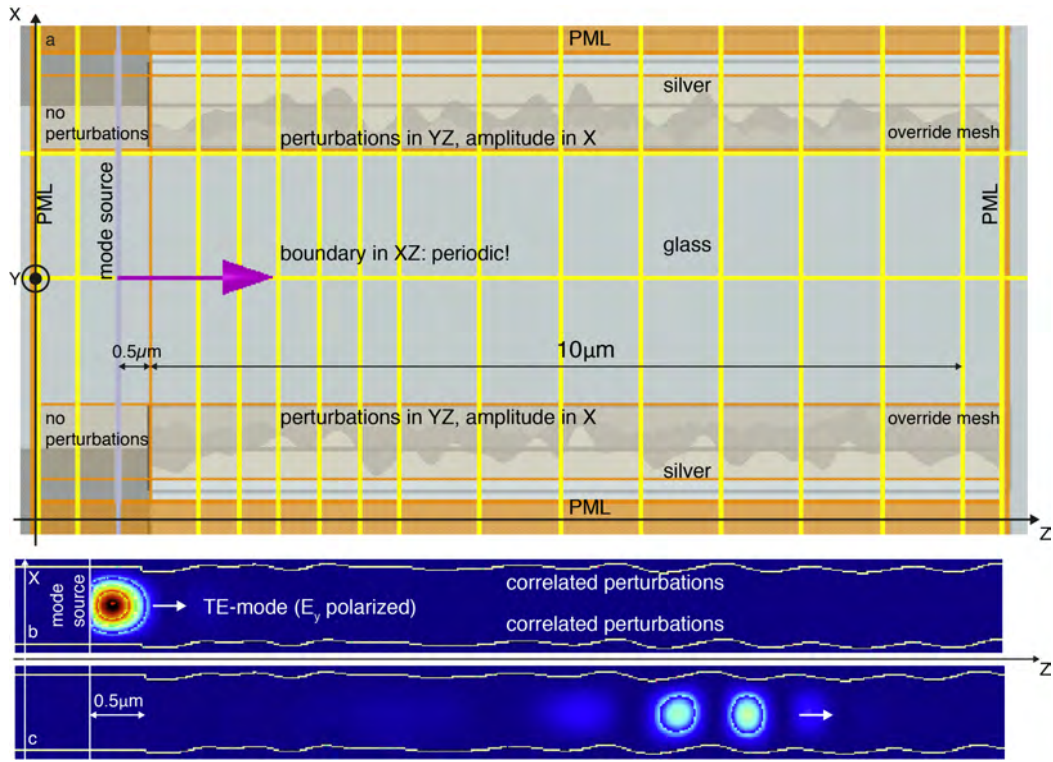

**Fig. S26: Sketch of the 3D FDTD simulation XZ-domain (Y=0) in Lumerical Solutions ($d$ = 535 nm). (a)** The mode source is positioned in the unperturbed waveguide with $d_{SiO2}$ = 536nm, exactly 0.5 $\mu$m before the start of the perturbation. **(b-c)** The intensity distribution of the TE-mode inside the perturbed ENZ waveguide in XZ with respect to the time for t ≈ 0 (b) and t ≈ t+Δt (c). In (b,c) the intensity |**E**|$^2$ is shown.

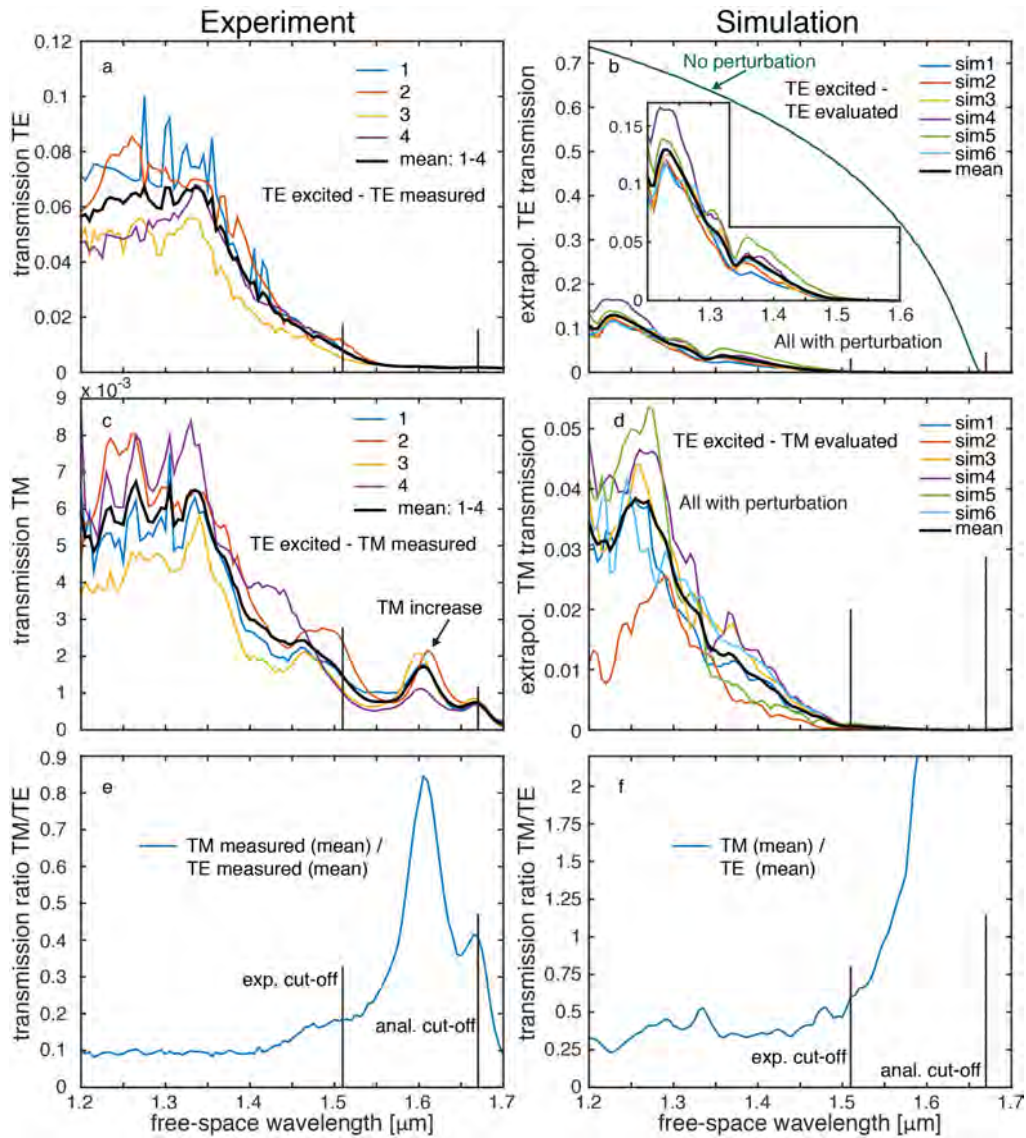

**Fig. S27: Experimental versus numerical, perturbed (3D FDTD, Lumerical Solutions®) transmission results ($d$ = 535 nm).** Excitation in the experiment and simulation: TE polarization. Experiment: four structures without objects, waveguide length = 33 $\mu$m. Simulation: six simulations, L x B x H = 10 $\mu$m x 8 $\mu$m x 0.535 $\mu$m, transmission exponentially extrapolated from $L$ = 10 $\mu$m to $L$ = 33 $\mu$m. **(a)** TE measurement with significant cut-off shift. **(c)** TM measurement with significant increase in TM polarization in between the experimental and analytical cut-off. **(e)** TM polarization rises from ca. 10 % to ca. 80 % in between the exp. cut-off and analytical cut-off. **(b)** Numeric TE detection with significant cut-off shift as in (a) and compared to transmission without perturbation. **(d)** Numeric TM detection. **(f)** Numeric TM polarization starts with ca. 30 % with respect to TE polarization and equalizes at $\lambda_0 \approx$ 1.56 $\mu$m. For larger wavelengths the TM amount rises vs. the TE amount.